\documentclass[preprint,prd,aps,showpacs,showkeys,nofootinbib]{revtex4}
\usepackage{graphicx}
\begin{document}

%\preprint{hep-ph/0412147}

\title{Electric dipole moments of neutron and heavy quarks in the B-LSSM}

\author{Jin-Lei Yang$^{1,2,3}$\footnote{yangjinlei@itp.ac.cn},
Tai-Fu Feng$^{1,2,4}$\footnote{fengtf@hbu.edu.cn}, Sheng-Kai Cui$^{1,2}$, Chang-xin Liu$^{1,2}$, Wei Li$^{1,2}$, Hai-Bin Zhang$^{1,2}$\footnote{hbzhang@hbu.edu.cn}}

\affiliation{Department of Physics, Hebei University, Baoding, 071002, China$^1$\\
Hebei Key Lab of High-precision Computation and Application of Quantum Field Theory, Baoding, 071002, China$^2$\\
CAS Key Laboratory of Theoretical Physics, School of Physical Sciences, University of Chinese Academy of Sciences, Beijing 100049, China$^3$\\
Department of Physics, Chongqing University, Chongqing 401331, China$^4$}

\begin{abstract}
The searching for the electric dipole moments (EDMs) of neutron ($d_n$), $b$ quark ($d_b$) and $c$ quark ($d_c$) gives strict upper bounds on these quantities. And recently, new upper bounds on $d_b$, $d_c$ are obtained by the strict limit on $d_n$. The models of new physics (NP) with additional CP-violating (CPV) sources are constrained strictly by these EDMs. In this work, we focus on the CPV effects on these EDMs in the minimal supersymmetric extension (MSSM) of the SM with local $B-L$ gauge symmetry (B-LSSM). The contributions from one-loop and some two-loop diagrams to the quark EDM are given in general form, which can also be used in the calculation of quark EDM in other models of NP. Considering the constrains from updated experimental data, the numerical results show that the two-loop corrections can make important contributions to these EDMs. Compared with the MSSM, the effects of new CPV phases and new parameters in the B-LSSM on these EDMs are also explored.

\end{abstract}


\maketitle

\section{Introduction\label{sec1}}
\indent\indent

In the standard model (SM), the only sources of CP-violating (CPV) are the Cabbibo-Kobayashi-Maskawa (CKM) phases, which appears to be the origin of the CPV phenomena observed in nondiagonal processes involving the $K$ and $B$ mesons~\cite{Christenson:1964fg,Abe:2001xe,Aubert:2002ic}. However, the observed baryon asymmetry of the universe indicates that the CPV sources in the SM are not sufficient, and new CPV sources are needed to generate the observed baryon asymmetry. In addition, it is well known that the theoretical predictions of electric dipole moments (EDMs) in the SM are too tiny to be detected in the near future. When new CPV phases are introduced, the theoretical predictions of EDMs can be enhanced vastly, hence the EDM of elementary particle is a clear signal of CPV~\cite{Ellis:1982tk,Polchinski:1983zd,Nath:1991dn,Kizukuri:1992nj,Kizukuri:1991mb}. As a result, studying the EDMs of elementary particles is of prime importance, and measurements on the EDMs of neutron or fundamental particles provide sensitive approaches to investigate potential new sources of CPV. So far no EDM for the neutron, $b$ quark or $c$ quark has been detected, but strong bounds on these quantities have been obtained~\cite{de-1,PDG,de-3,Blinov:2008mu,Sala:2013osa}
\begin{eqnarray}
&&|d_n|<3.0\times10^{-26}{\rm e\cdot cm},\nonumber\\
&&|d_b|<2.0\times10^{-17}{\rm e\cdot cm},\nonumber\\
&&|d_c|<4.4\times10^{-17}{\rm e\cdot cm}.
\label{EDMlimits}
\end{eqnarray}
In addition, the chromo-EDM (CEDM) of heavy quark is constrained by the strict limit on the neutron EDM. The EDM of neutron can be expressed in terms of fundamental dipoles~\cite{Pospelov:2000bw}
\begin{eqnarray}
&&d_n=(1\pm0.5)[1.4(d_d^\gamma-0.25d_u^\gamma)+1.1e(d_d^g+0.5d_u^g)]\pm(22\pm10){\;\rm MeV}C_5,
\label{EDMdn}
\end{eqnarray}
where $d_q^\gamma$, $d_q^g$, $C_5$ denote the quark EDM of $q$ from the electroweak interaction, the CEDM of $q$ and the coefficient of Weinberg operator at the chirality scale respectively. Using the running from Refs.~\cite{Braaten:1990gq,Degrassi:2005zd} at one-loop, $d_{d,u}^{\gamma,g}$ and $C_5$ can be expressed in terms of $d_c^g$ at the scale $m_c$~\cite{Sala:2013osa}. Combining with the limits on $d_n$, new upper limit on $d_c^g$ is obtained~\cite{Sala:2013osa}, $|d_c^g|<1.0\times10^{-22}{\rm cm}$. Then assuming constructive interference between the EDM and CEDM contributions at the NP scale~\cite{Gisbert:2019ftm}, new bounds on the EDM of $b$ and $c$ quark are derived by using the stringent limits on $d_b^g$ in Ref.~\cite{Chang:1990jv} and $d_c^g$ in Ref.~\cite{Sala:2013osa}
\begin{eqnarray}
&&|d_b|<1.2\times10^{-20}{\rm e\cdot cm},\nonumber\\
&&|d_c|<1.5\times10^{-21}{\rm e\cdot cm},
\end{eqnarray}
which improves the previous ones in Eq.~(\ref{EDMlimits}) by about three orders of magnitude. Since the experimental upper bounds on these quantities are very small, the contributions from new CPV phases are limited strictly by the present experimental data, hence researching NP effects on these EDMs may shed light on the mechanism of CPV.

In extensions of the SM, the supersymmetry is considered as one of the most plausible candidates. For the explanation of baryon asymmetry, electroweak baryogenesis (EWB) is one of the most well-known mechanisms, and new CPV phases are needed to enhance the asymmetry in this case. In Refs.~\cite{Dine:1990fj,Cohen:1992zx,Huet:1995sh,Lee:2004we,Konstandin:2005cd}, EWB is discussed in the MSSM in great detail, and the results show that the $\mu$ term (the bilinear Higgs mass term in the superpotential) is the dominant source of baryon asymmetry. However, when the phase of $\mu$ is taken to be large, the theoretical predictions of EDMs are several magnitudes larger than the corresponding upper bounds. The effects have been explored in Refs.~\cite{Falk:1996ni,Falk:1998pu,Brhlik:1998zn,Bartl:1999bc,Abel:2001vy,Barger:2001nu,
Olive:2005ru,Cirigliano:2006dg,YaserAyazi:2006zw}. The results show that the most interesting possibility to suppress these EDMs to below the corresponding experimental upper bounds is the contributions from different phases cancel each other. The CPV characters in supersymmetry are very interesting and studies on them may shed some light on the general characteristics of the supersymmetric model.

In this work, we explore the CPV effects on the EDM of neutron $d_n$, $b$ quark $d_b$ and $c$ quark $d_c$ in the MSSM with local $B-L$ gauge symmetry (B-LSSM)~\cite{Barger:2008wn,FileviezPerez:2008sx,5,6}. The model is based on the gauge symmetry group $SU(3)\otimes SU(2)_L\otimes U(1)_Y\otimes U(1)_{B-L}$, where $B$ stands for the baryon number and $L$ stand for the lepton number respectively. Compared with the MSSM, there are much more candidates for the dark matter~\cite{16,1616,DelleRose:2017ukx,DelleRose:2017uas} in the B-LSSM, which also accounts elegantly for the existence and smallness of the left-handed neutrino masses. Since the exotic singlet Higgs and right-handed (s) neutrinos ~\cite{77,88,9,99,10,11} releases additional parameter space from the LHC constraints, the model alleviates the little hierarchy problem of the MSSM~\cite{search}. In addition, the invariance under $U(1)_{B-L}$ gauge group imposes the R-parity conservation, which is assumed in the MSSM to avoid proton decay. And R-parity conservation can be maintained if $U(1)_{B-L}$ symmetry is broken spontaneously~\cite{C.S.A}.

The paper is organized as follows. In Sec. II, the main ingredients of the B-LSSM are summarized briefly by
introducing the superpotential and the general soft breaking terms. Then the analysis on the EDM of neutron $d_n$, $b$ quark $d_b$ and $c$ quark $d_c$ are presented in Sec. III. In order to see the corrections to these EDMs clearly, the numerical results of $d_n$, $d_b$, $d_c$ with new CPV phases are explored in Sec. IV. Conclusions are summarized in Sec. V. Some relevant mass matrixes and interaction vertexes are collected in appendix A and appendix B respectively.

\section{The B-LSSM\label{sec2}}

Besides the superfields of the MSSM, two chiral singlet superfields $\hat{\eta}_{1}\sim(1,1,0,-1)$, $\hat{\eta}_{2}\sim(1,1,0,1)$ and three generations of right-handed neutrinos are introduced in the B-LSSM. And the local gauge group is enlarged to $SU(3)_C\otimes SU(2)_L\otimes U(1)_Y\otimes U(1)_{B-L}$ in the model, where the $U(1)_{B-L}$ is spontaneously broken by the chiral singlets. This version of B-LSSM is encoded in SARAH~\cite{164}, which is used to create the mass matrices and interaction vertexes in the model. Then the local gauge symmetry $SU(2)_L\otimes U(1)_Y\otimes U(1)_{B-L}$ breaks down to the electromagnetic symmetry $U(1)_{em}$ as the Higgs fields receive vacuum expectation values:
\begin{eqnarray}
&&H_1^1=\frac{1}{\sqrt2}(v_1+{\rm Re}H_1^1+i{\rm Im}H_1^1),
\qquad\; H_2^2=\frac{1}{\sqrt2}(v_2+{\rm Re}H_2^2+i{\rm Im}H_2^2),\nonumber\\
&&\tilde{\eta}_1=\frac{1}{\sqrt2}(u_1+{\rm Re}\tilde{\eta}_1+i{\rm Im}\tilde{\eta}_1),
\qquad\;\quad\;\tilde{\eta}_2=\frac{1}{\sqrt2}(u_2+i{\rm Re}\tilde{\eta}_2+i{\rm Im}\tilde{\eta}_2)\;.
\end{eqnarray}
Then in analogy to the ratio of the MSSM VEVs ($\tan\beta=\frac{v_2}{v_1}$), we can define $\tan\beta^{'}=\frac{u_2}{u_1}$.

In addition, the superpotential of the B-LSSM can be written as
\begin{eqnarray}
&&W=Y_u^{ij}\hat{Q_i}\hat{H_2}\hat{U_j^c}+\mu \hat{H_1} \hat{H_2}-Y_d^{ij} \hat{Q_i} \hat{H_1} \hat{D_j^c}
-Y_e^{ij} \hat{L_i} \hat{H_1} \hat{E_j^c}+\nonumber\\
&&\;\;\;\;\;\;\;\;\;Y_{\nu, ij}\hat{L_i}\hat{H_2}\hat{\nu}^c_j-\mu' \hat{\eta}_1 \hat{\eta}_2
+Y_{x, ij} \hat{\nu}_i^c \hat{\eta}_1 \hat{\nu}_j^c,
\end{eqnarray}
where $i, j$ are generation indices. Correspondingly, the soft breaking terms of the B-LSSM are generally given as
\begin{eqnarray}
&&\mathcal{L}_{soft}=\Big[-\frac{1}{2}(M_1\tilde{\lambda}_{B} \tilde{\lambda}_{B}+M_2\tilde{\lambda}_{W} \tilde{\lambda}_{W}+M_3\tilde{\lambda}_{g} \tilde{\lambda}_{g}+2M_{BB'}\tilde{\lambda}_{B'} \tilde{\lambda}_{B}+M_{B'}\tilde{\lambda}_{B'} \tilde{\lambda}_{B'})-
\nonumber\\
&&\hspace{1.4cm}
B_\mu H_1H_2 -B_{\mu'}\tilde{\eta}_1 \tilde{\eta}_2 +T_{u,ij}\tilde{Q}_i\tilde{u}_j^cH_2+T_{d,ij}\tilde{Q}_i\tilde{d}_j^cH_1+
T_{e,ij}\tilde{L}_i\tilde{e}_j^cH_1+T_{\nu}^{ij} H_2 \tilde{\nu}_i^c \tilde{L}_j+\nonumber\\
&&\hspace{1.4cm}
T_x^{ij} \tilde{\eta}_1 \tilde{\nu}_i^c \tilde{\nu}_j^c+h.c.\Big]-m_{\tilde{\nu},ij}^2(\tilde{\nu}_i^c)^* \tilde{\nu}_j^c-
m_{\tilde{q},ij}^2\tilde{Q}_i^*\tilde{Q}_j-m_{\tilde{u},ij}^2(\tilde{u}_i^c)^*\tilde{u}_j^c-m_{\tilde{\eta}_1}^2 |\tilde{\eta}_1|^2-\nonumber\\
&&\hspace{1.4cm}
m_{\tilde{\eta}_2}^2 |\tilde{\eta}_2|^2-m_{\tilde{d},ij}^2(\tilde{d}_i^c)^*\tilde{d}_j^c-m_{\tilde{L},ij}^2\tilde{L}_i^*\tilde{L}_j-
m_{\tilde{e},ij}^2(\tilde{e}_i^c)^*\tilde{e}_j^c-m_{H_1}^2|H_1|^2-m_{H_2}^2|H_2|^2,
\end{eqnarray}
where $\tilde\lambda_{B}, \tilde\lambda_{B'}$ denote the gauginos of $U(1)_Y$ and $U(1)_{(B-L)}$ respectively.

It can be noted that there are two Abelian groups in the B-LSSM, which gives rise to a new effect absent in the MSSM or other SUSY models with just one Abelian gauge group: the gauge kinetic mixing. This mixing couples the $B-L$ sector to the MSSM sector, and it can be induced through RGEs~\cite{RGE1,RGE2,RGE3,RGE4,RGE5,RGE6,RGE7} even if it is set to zero at $M_{GUT}$. Immediate interesting consequences of the gauge kinetic mixing arise in various sectors of the model. Firstly, the gauge kinetic mixing leads to the mixing between the $H_1^1,\;H_2^2,\;\tilde{\eta}_1,\;\tilde{\eta}_2$ at the tree level, which changes the vacuum structure vastly and also affects the theoretical prediction of the SM-like Higgs boson mass. Meanwhile, $\tilde\lambda_{B'}$ mixes with the two higgsinos in the MSSM at the tree level. Then new gauge boson $Z'$ mixes with the $Z$ boson in the MSSM, and new gauge coupling constant $g_{_{YB}}$ is introduced. In addition, additional D-terms contribute to the mass matrices of the squarks and sleptons which affects the theoretical predictions of various observations of the model. All of these properties of the model are introduced in detail in our earlier work~\cite{Yang:2018utw,Yang:2018guw}.

Comparing with the MSSM, new definitions of some mass matrixes appearing in the calculation of EDMs are collected in appendix \ref{mass matrix} in order to see new contributions to EDMs in the B-LSSM clearly. Squarks play an important role in the theoretical predictions of EDMs. Comparing with the MSSM, it can be noted in Eq.~(\ref{massd}), Eq.~ (\ref{massu}) that new gauge coupling constants $g_{_B}$, $g_{_{YB}}$ and VEVs $u_1$, $u_2$ can affect the mass matrixes of both up and down type squarks. As a result, different theoretical predictions of EDMs are produced in B-LSSM compared with the ones in MSSM. In addition, new neutralinos which can also make contributions to EDMs are introduced in the B-LSSM, and the corresponding mass matrix is collected in appendix \ref{mass matrix}.

\section{The EDMs of neutron and heavy quarks\label{sec3}}

For the neutron EMD $d_n$, we adopt the values $0.5$ and $12{\;\rm MeV}$ for the coefficients $1\pm0.5$ and $22\pm10{\;\rm MeV}$ in Eq.~(\ref{EDMdn}) respectively, in order to coincide with the discussion in Ref.~\cite{Sala:2013osa}. In addition, the quark EDM at the low scale can be written as
\begin{eqnarray}
&&d_q=d_q^\gamma(\Lambda_\chi)+\frac{e}{4\pi}d_q^g(\Lambda_\chi)+\frac{e\Lambda_\chi}{4\pi}C_5(\Lambda_\chi),
\end{eqnarray}
where $\Lambda_\chi=m_q$ denotes the chirality breaking scale, $m_q$ denotes the corresponding quark mass. The Wilson coefficient of the purely gluonic Weinberg operator originates from the two-loop "gluino-squark" diagrams, and the concrete expression of $C_5$ can be written as~\cite{Dai:1990xh,Dicus:1989va,Ibrahim:1998je}
\begin{eqnarray}
&&C_5(\Lambda)=-\frac{3g_3^5}{(4\pi)^4M_3^3}\Big\{m_t \Im[e^{2i\theta_3}(Z_{\tilde t})_{2,2}(Z_{\tilde t})_{2,1}^\dagger]\frac{x_{\tilde t_1}-x_{\tilde t_2}}{x_{M_3}}H\Big(\frac{x_{\tilde t_1}}{x_{M_3}},\frac{x_{\tilde t_2}}{x_{M_3}},\frac{x_t}{x_{M_3}}\Big)\nonumber\\
&&\qquad\qquad\;\;+m_b \Im[e^{2i\theta_3}(Z_{\tilde b})_{2,2}(Z_{\tilde b})_{2,1}^\dagger]\frac{x_{\tilde b_1}-x_{\tilde b_2}}{x_{M_3}}H\Big(\frac{x_{\tilde b_1}}{x_{M_3}},\frac{x_{\tilde b_2}}{x_{M_3}},\frac{x_b}{x_{M_3}}\Big)\Big\},
\end{eqnarray}
where $\Lambda$ denotes the matching scale, $Z_{\tilde t}(Z_{\tilde b})$ is the diagonalizing matrix for the squared mass matrix of stop (sbottom), and the function $H$ can be found in Refs.~\cite{Dai:1990xh,Dicus:1989va,Ibrahim:1998je}.

Meanwhile, $d_q^\gamma$, $d_q^g$ and $C_5$ are evolved with the renormalization group equations from the matching scale $\Lambda$ down to the chirality breaking scale $\Lambda_\chi$~\cite{Arnowitt:1990eh,Dekens:2013zca,Chien:2015xha,Yamanaka:2017mef} according to
\begin{eqnarray}
&&d_q^\gamma(\Lambda_\chi)=1/1.53d_q^\gamma(\Lambda),\;d_q^g(\Lambda_\chi)=3.4d_q^g(\Lambda),\;
C_5(\Lambda_\chi)=3.4C_5(\Lambda).
\end{eqnarray}

The effective Lagrangian for the quark EDMs can be written as
\begin{eqnarray}
&&\mathcal{L}_{EDM}=-\frac{i}{2}d_q^\gamma\bar q \sigma^{\mu\nu}\gamma_5 q F_{\mu\nu},
\end{eqnarray}
where $\sigma^{\mu\nu}=i[\gamma^\mu,\gamma^\nu]/2$, $q$ is the wave function for quark, and $F_{\mu\nu}$ is the electromagnetic field strength. Adopting the effective Lagrangian approach, the quark EDMs can be written as
\begin{eqnarray}
&&d_q^\gamma=-\frac{2e Q_q m_q}{(4\pi)^2}\Im(C_2^R+C_2^{L*}+C_6^R),
\label{EDM}
\end{eqnarray}
where $C_{2,6}^{L,R}$ represent the Wilson coefficients of the corresponding operators $O_{2,6}^{L,R}$
\begin{eqnarray}
&&O_2^{L,R}=\frac{e Q_q}{(4\pi)^2}(-iD_\alpha^*) \bar q \gamma^\alpha F\cdot\sigma P_{L,R}q,\nonumber\\
&&O_6^{L,R}=\frac{e Q_q m_q}{(4\pi)^2}\bar q F\cdot\sigma P_{L,R}q.
\end{eqnarray}

Similarly, the effective Lagrangian for the quark CEDMs can be written as
\begin{eqnarray}
&&\mathcal{L}_{CEDM}=-\frac{i}{2} d_q^g \bar q\sigma^{\mu\nu}\gamma_5 q G_{\mu\nu}^a T^a,
\end{eqnarray}
where $G_{\mu\nu}$ is the $SU(3)$ gauge field strength, $T^a$ is the $SU(3)$ generators. Then the quark CEDMs can be written as
\begin{eqnarray}
&&d_q^g=-\frac{2g_3m_q}{(4\pi)^2}\Im(C_7^R+C_7^{L*}+C_8^R),
\label{CEDM}
\end{eqnarray}
where $C_{7,8}^{L,R}$ represent the Wilson coefficients of the corresponding operators $O_{7,8}^{L,R}$
\begin{eqnarray}
&&O_7^{L,R}=\frac{g_3}{(4\pi)^2}(-iD_\alpha^*) \bar q \gamma^\alpha G^a\cdot\sigma T^a P_{L,R}q,\nonumber\\
&&O_8^{L,R}=\frac{g_3m_q}{(4\pi)^2}\bar q G^a\cdot\sigma T^a P_{L,R}q.
\end{eqnarray}

\begin{figure}
\setlength{\unitlength}{1mm}
\centering
\includegraphics[width=6.4in]{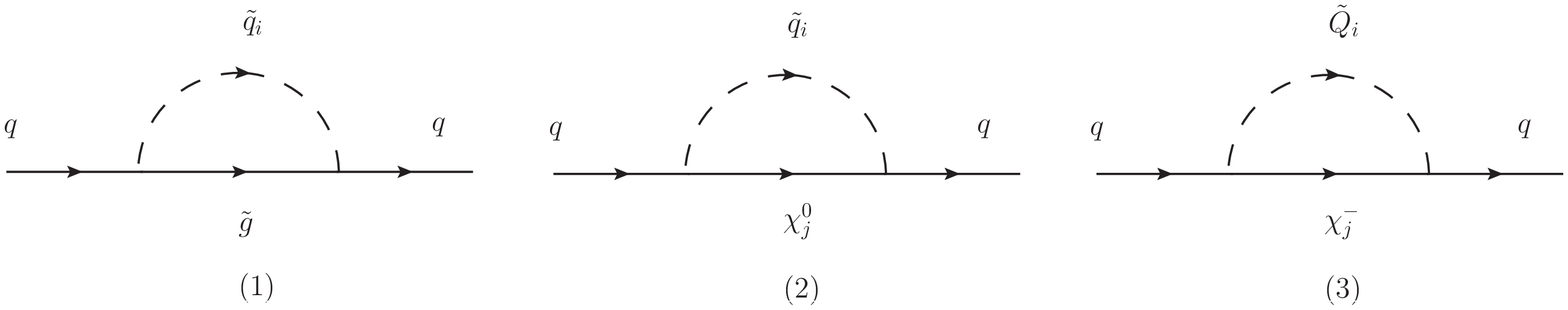}
\vspace{0cm}
\caption[]{The one-loop diagrams which contributes to $d_q^\gamma$ and $d_q^g$ are obtained by attaching a photon and a gluon respectively to the internal particles in all possible ways.}
\label{Feynman diagram Q}
\end{figure}

The one-loop Feynman diagrams contributing to the above amplitudes are depicted in Fig.~\ref{Feynman diagram Q}.
Calculating the Feynman diagrams, $d_q^\gamma$ and $d_q^g$ at the one-loop level can be written as
\begin{eqnarray}
&&d_q^{\gamma(1)}=\frac{e_q e}{12\pi^2m_W}\frac{\sqrt {x_{\tilde g}}}{x_{\tilde q_i}}\Im\Big[C_{\bar{\tilde g} \tilde q_i q}^L C_{\bar q \tilde q_i \tilde g}^L\Big]I_1\Big(\frac{x_{\tilde g}}{x_{\tilde q_i}}\Big),\nonumber\\
&&d_q^{g(1)}=\frac{-g_3}{32\pi^2m_W}\frac{\sqrt {x_{\tilde g}}}{x_{\tilde q_i}}\Im\Big[C_{\bar {\tilde g} \tilde q_i q}^L C_{\bar q \tilde q_i \tilde g}^L\Big]I_2\Big(\frac{x_{\tilde g}}{x_{\tilde q_i}}\Big),\nonumber\\
&&d_q^{\gamma(2)}=\frac{e_q e}{32\pi^2m_W}\frac{\sqrt {x_{\chi_j^0}}}{x_{\tilde q_i}}\Im\Big[C_{\bar q \tilde q_i\chi_j^0}^L C_{\bar{\chi}_j^0 \tilde q_i q}^R\Big]I_1\Big(\frac{x_{\chi_j^0}}{x_{\tilde q_i}}\Big),\nonumber\\
&&d_q^{g(2)}=\frac{g_3^3}{128\pi^2 e^2 m_W}\frac{\sqrt {x_{\chi_j^0}}}{x_{\tilde q_i}}\Im\Big[C_{\bar q \tilde q_i\chi_j^0}^L C_{\bar{\chi}_j^0 \tilde q_i q}^R\Big]I_1\Big(\frac{x_{\chi_j^0}}{x_{\tilde q_i}}\Big),\nonumber\\
&&d_q^{\gamma(3)}=\frac{e}{16\pi^2m_W}\frac{\sqrt {x_{\chi_j^-}}}{x_{\tilde Q_i}}\Im\Big[C_{\bar q \tilde Q_i\chi_j^-}^L C_{\bar{\chi}_j^- \tilde Q_i q}^R\Big]\Big[e_QI_1\Big(\frac{x_{\chi_j^-}}{x_{\tilde Q_i}}\Big)+(e_q-e_Q)I_3\Big(\frac{x_{\chi_j^-}}{x_{\tilde q_i}}\Big)\Big],\nonumber\\
&&d_q^{g(3)}=\frac{g_3^3}{16\pi^2e^2m_W}\frac{\sqrt {x_{\chi_j^-}}}{x_{\tilde Q_i}}\Im\Big[C_{\bar q \tilde Q_i\chi_j^-}^L C_{\bar{\chi}_j^- \tilde Q_i q}^R\Big]I_1\Big(\frac{x_{\chi_j^-}}{x_{\tilde Q_i}}\Big),
\label{17}
\end{eqnarray}
where $x_i$ denotes $m_i^2/m_W^2$, $g_3$ is the strong coupling constant, $C_{abc}^{L,R}$ denotes the constant parts of the interaction vertex about $abc$ which can be obtained through SARAH, and $a$, $b$, $c$ denote the interacting particles. In order to see the contributions to EDMs introduced by the B-LSSM in addition to those already present in the MSSM, all constants $C_{abc}^{L,R}$ appeared in our calculation are collected in appendix \ref{constants}. The functions $I_{1,2,3}$ can be written as
\begin{eqnarray}
&&I_1(x)=\frac{1}{2(x-1)^2}(1+x+\frac{2x}{x-1}\ln x),\\
&&I_2(x)=\frac{1}{6(x-1)^2}(10x-26-\frac{2x-18}{x-1}\ln x),\\
&&I_3(x)=\frac{1}{2(x-1)^2}(3-x+\frac{2}{x-1}\ln x).
\end{eqnarray}

\begin{figure}
\setlength{\unitlength}{1mm}
\centering
\includegraphics[width=6in]{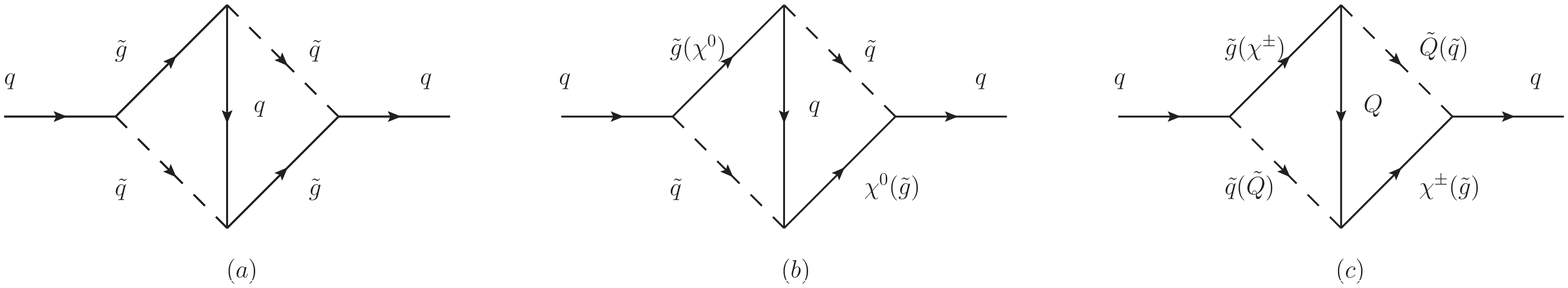}
\vspace{0cm}
\caption[]{The two-loop diagrams which contributes to $d_q^\gamma$ and $d_q^g$ are obtained by attaching a photon and a gluon respectively to the internal particles in all possible ways.}
\label{two-loop Feynman diagram Q}
\end{figure}

The two-loop gluino corrections to the Wilson coefficients from the self-energy diagrams for quarks are considered, the corresponding Feynman diagrams are depicted in Fig.~\ref{two-loop Feynman diagram Q}. The corresponding diploe moment diagrams are obtained by attaching a photon or gluon to the internal particles in all possible ways.
Then the contributions from these two-loop diagrams to $d_q^\gamma$ and $d_q^g$ can be written as
\begin{eqnarray}
&&d_q^{\gamma(a)}=\frac{-4e_q e g_3^2|m_{\tilde g}|}{9(4\pi)^4m_W^2}F_3(x_{_q},x_{_{\tilde q_j}},x_{_{\tilde g}},x_{_{\tilde g}},x_{_{\tilde q_i}})\Im[C_{\bar q \tilde g \tilde q_j}^L C_{\bar {\tilde g}q\tilde q_j}^L],\nonumber\\
&&d_q^{g(a)}=d_q^{\gamma(a)}g_3/(e_q e),\nonumber\\
&&d_q^{\gamma(b)}=\frac{4 e_q e}{3(4\pi)^4m_W^2}\Big\{|m_{\tilde g}|F_4(x_{_q},x_{_{\tilde q_j}},x_{_{\tilde g}},x_{_{\chi_k^0}},x_{_{\tilde q_i}})\Im[C_{\bar \chi_k^0 q \tilde q_j}^RC_{\bar \chi_k^0 q \tilde q_i}^L C_{\bar {\tilde g}q\tilde q_j}^LC_{\bar {\tilde g}q\tilde q_i}^L-C_{\bar \chi_k^0 q \tilde q_j}^L\nonumber\\
&&\qquad\quad \times C_{\bar \chi_k^0 q \tilde q_i}^RC_{\bar q \tilde g \tilde q_j}^{L*}C_{\bar q \tilde g \tilde q_i}^{L*}]-m_{\chi_k^0}F_5(x_{_q},x_{_{\tilde q_j}},x_{_{\tilde g}},x_{_{\chi_k^0}},x_{_{\tilde q_i}})\Im[C_{\bar \chi_k^0 q \tilde q_j}^RC_{\bar \chi_k^0 q \tilde q_i}^RC_{\bar q \tilde g \tilde q_j}^{L*}C_{\bar {\tilde g}q\tilde q_i}^L\nonumber\\
&&\qquad\quad-C_{\bar \chi_k^0 q \tilde q_j}^LC_{\bar \chi_k^0 q \tilde q_i}^LC_{\bar q \tilde g \tilde q_i}^{L*}C_{\bar {\tilde g}q\tilde q_j}^L]\Big\},\nonumber\\
&&d_q^{g(b)}=d_q^{\gamma(b)}g_3/(e_q e),\nonumber\\
&&d_q^{\gamma(c)}=\frac{2 e}{3(4\pi)^4m_W^2}\Big\{|m_{\tilde g}|F_4(x_{_Q},x_{_{\tilde Q_j}},x_{_{\tilde g}},x_{_{\chi_k^\pm}},x_{_{\tilde q_i}})\Im[C_{\bar Q \chi_k^\pm \tilde q_j}^LC_{\bar q\chi_k^\pm \tilde Q_i}^RC_{\bar{\tilde g}Q\tilde Q_j}^LC_{\bar{\tilde g}q\tilde q_i}^L-C_{\bar Q \chi_k^\pm \tilde q_j}^R\nonumber\\
&&\qquad\quad\times C_{\bar q\chi_k^\pm \tilde Q_i}^LC_{\bar Q\tilde g\tilde Q_j}^{L*}C_{\bar q\tilde g\tilde q_i}^{L*}]-m_{\chi_k^\pm}F_5(x_{_Q},x_{_{\tilde Q_j}},x_{_{\tilde g}},x_{_{\chi_k^\pm}},x_{_{\tilde q_i}})\Im[C_{\bar Q \chi_k^\pm \tilde q_j}^LC_{\bar q\chi_k^\pm \tilde Q_i}^LC_{\bar Q\tilde g\tilde Q_j}^{L*}C_{\bar{\tilde g}q\tilde q_i}^L\nonumber\\
&&\qquad\quad-C_{\bar Q \chi_k^\pm \tilde q_j}^RC_{\bar q\chi_k^\pm \tilde Q_i}^RC_{\bar{\tilde g}Q\tilde Q_j}^LC_{\bar q\tilde g\tilde q_i}^{L*}]\Big\},\nonumber\\
&&d_q^{g(c)}=d_q^{\gamma(c)}g_3/e,
\label{21}
\end{eqnarray}
where the concrete expressions for the functions $F_{3,4,5}$ can be found in Ref.~\cite{Feng:2004vu}. The similar expressions of two-loop gluino contributions can be found in Ref.~\cite{Feng:2004vu}. We translate them into our notations which is written in general form and can be used in the calculation of quark EDM in other models of NP.

\begin{figure}
\setlength{\unitlength}{1mm}
\centering
\includegraphics[width=4in]{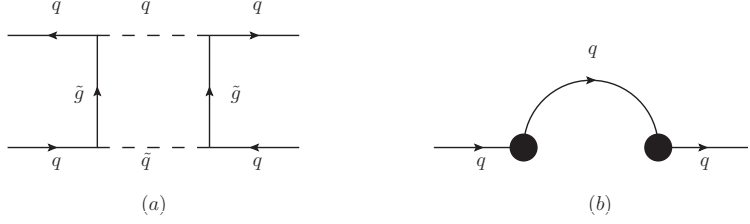}
\vspace{0cm}
\caption[]{Full theory diagram (a) and effective diagram (b) are plotted, where the blobs denote the effective vertexes, and a outgoing photon or gluon is attached by all possible ways.}
\label{matching Feynman diagram}
\end{figure}

We should note that there are infrared divergencies in Fig.~\ref{two-loop Feynman diagram Q} when the SM quarks appear as internal particles, because we calculate these diagrams by expanding the external momentum. In this case, matching full theory diagrams to the corresponding two-loop diagrams in Fig.~\ref{two-loop Feynman diagram Q} is needed to cancel the infrared divergency. Taking Fig.~\ref{two-loop Feynman diagram Q}(a) as example to illustrate how to cancel the infrared divergency, the corresponding diagrams are shown in Fig.~\ref{matching Feynman diagram}.
When the external gluon is attached to an internal particle in Fig.~\ref{two-loop Feynman diagram Q}(a), and the external gluon can be attached to the same internal particle in Fig.~\ref{matching Feynman diagram}(a) or (b). Then infrared divergency in the diagram by attaching a gluon in Fig.~\ref{two-loop Feynman diagram Q}(a) can be cancelled by subtracting the corresponding diagram by attaching a gluon in the same way in Fig.~\ref{matching Feynman diagram}.

\begin{figure}
\setlength{\unitlength}{1mm}
\centering
\includegraphics[width=6in]{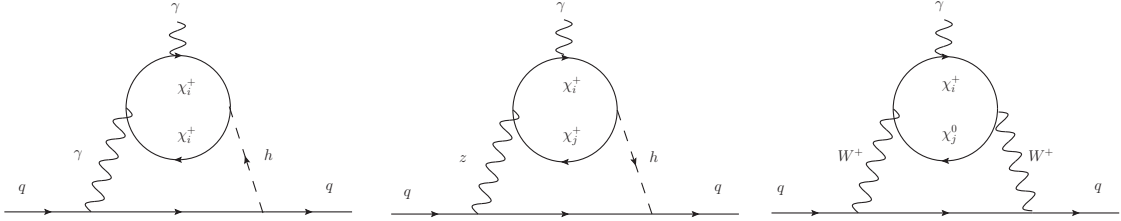}
\vspace{0cm}
\caption[]{The two-loop Barr-Zee type diagrams contributing to the quark EDM. The diagrams in which the photon or gluon is emitted from the $W$ boson or the internal fermion do not contribute to the quark EDM or CEDM.}
\label{two-loop Feynman Bazz}
\end{figure}

In addition, the two-loop Barr-Zee type diagrams can also make contributions to the quark EDM. The diagrams in which a closed fermion loop is attached to the virtual gauge bosons or Higgs fields are considered, and the corresponding Feynman diagrams are depicted in Fig.~\ref{two-loop Feynman Bazz}.
Then, the contributions from these two-loop Barr-Zee type diagrams to $d_q^\gamma$ are given by~\cite{Giudice:2005rz}
\begin{eqnarray}
&&d_q^{\gamma h}=\frac{e_q e^3}{32\pi^4 m_W}\frac{\sqrt{x_{\chi_i^+}}}{x_{h_k}}\Im\Big[C_{\bar\chi_i^+ h \chi_i^+}^R C_{\bar q h_k q}\Big]f_{\gamma H}\Big(\frac{x_{\chi_i^+}}{x_{h_k}}\Big),\nonumber\\
&&d_q^{Zh}=\frac{e^2(T_{3q}-2e_q s_w^2)}{128\pi^4 c_w s_w m_W}\frac{\sqrt{x_{\chi_i^+}}}{x_{h_k}}\Im\Big[\Big(C_{\bar\chi_j^+ h_k \chi_i^+}^RC_{\bar\chi_i^+ Z \chi_j^+}^L-C_{\bar\chi_j^+ h_k \chi_i^+}^LC_{\bar\chi_i^+ Z \chi_j^+}^R\Big)C_{\bar q h_k q}\Big]\nonumber\\
&&\qquad\;\;\;*f_{ZH}\Big(\frac{x_Z}{x_{h_k}},\frac{x_{\chi_i^+}}{x_{h_k}},\frac{x_{\chi_j^+}}{x_{h_k}}\Big),\nonumber\\
&&d_q^{WW}=\frac{T_{3q} e^3}{128\pi^4 s_w^2 m_W}\sqrt{x_q x_{\chi_i^+}x_{\chi_j^0}}\Im\Big[C_{\bar\chi_j^0 W^+_\mu \chi_i^+}^RC_{\bar\chi_j^0 W^+_\mu \chi_i^+}^{L*}\Big]f_{WW}\Big(x_{\chi_i^+},x_{\chi_j^0}\Big),
\label{22}
\end{eqnarray}
where $s_w\equiv\sin \theta_W,\;c_w\equiv\cos \theta_W$, and $\theta_W$ is the Weinberg angle, $T_{3q}$ denotes the isospin of the corresponding quark, the functions $f_{\gamma H},\;f_{ZH},\;f_{WW}$ can be found in Ref.~\cite{Giudice:2005rz}.

%%%%%%%%%%%%%%%%%%%%%%%%%%%%%END REVISED%%%%%%%%%%%%%%%%%%%%%%%%%%%%%%%%%%%%%

\section{Numerical analysis\label{sec4}}

In this section, we present the numerical results of the EDMs $d_n$, $d_b$ and $d_{c}$ in the B-LSSM. For SM parameters, we take $m_W=80.385{\;\rm GeV},\;m_Z=90.1876{\;\rm GeV},\;m_u=2.3{\;\rm MeV},\;m_d=4.8{\;\rm MeV},\;m_b=4.65{\;\rm GeV},\;m_c=1.275{\;\rm GeV},\;\alpha_{em}(m_Z)=1/128.9,\;\alpha_s(m_Z)=0.118$. The SM-like Higgs boson mass is $125.09 {\;\rm GeV}$~\cite{Aad:2015zhl}. The updated experimental data~\cite{newZ} on searching $Z'$ indicates $M_{Z'}\geq4.05{\;\rm TeV}$ at $95\%$ Confidence Level (CL). And the analysis of the data collected at the large electron-positron (LEP) collider about 4-fermion's interactions gives us an upper bound on the ratio between the $Z'$ mass and its gauge coupling at $99\%$ CL as $M_{Z'}/g_B>6{\;\rm TeV}$~\cite{20,21}. Experimental data collected at LEP also have been used to obtain upper bound on the $Z-Z'$ mixing angle $\theta_{W'}$~\cite{Barate:1999qx,Abreu:2000ap}. The bound is model-dependent, but current data from collider yields $\sin \theta_{W'}\lesssim10^{-3}$ for standard GUT models. The upper bound on $Z-Z'$ mixing angle is considered in the following analysis. In addition, the LHC experimental data also constrain $\tan\beta'<1.5$. Since the contributions from heavy $Z'$ are highly suppressed, we approximately fix $M_{Z'}=4.2{\;\rm TeV}$ without losing generality. The constraints on the parameter space of the B-LSSM from the experimental data on $\bar B\rightarrow X_s\gamma$, $B_s^0\rightarrow \mu^+\mu^-$ are studied in Ref.~\cite{JLYang:2018}, the results show that the charged Higgs boson mass are limited in the range $M_{H^\pm}\gtrsim1 {\;\rm TeV}$. Hence the contributions of charged Higgs boson are highly suppressed in this case, and we can fix $M_{H^\pm}=1.5 {\;\rm TeV}$ approximately without losing generality. All parameters fixed above affect the numerical results negligibly. For the squark sector, we take $m_{\tilde q}=m_{\tilde u}=m_{\tilde d}=diag(M_Q,M_Q,M_Q){\;\rm TeV}$ and $T_{u,d}=Y_{u,d}\;diag(A_Q,A_Q,A_Q) {\;\rm TeV}$ for simplicity. The observed Higgs signal limits that $M_Q>1.5$~\cite{Un:2016hji}.

\begin{figure}
\setlength{\unitlength}{1mm}
\centering
\includegraphics[width=2.7in]{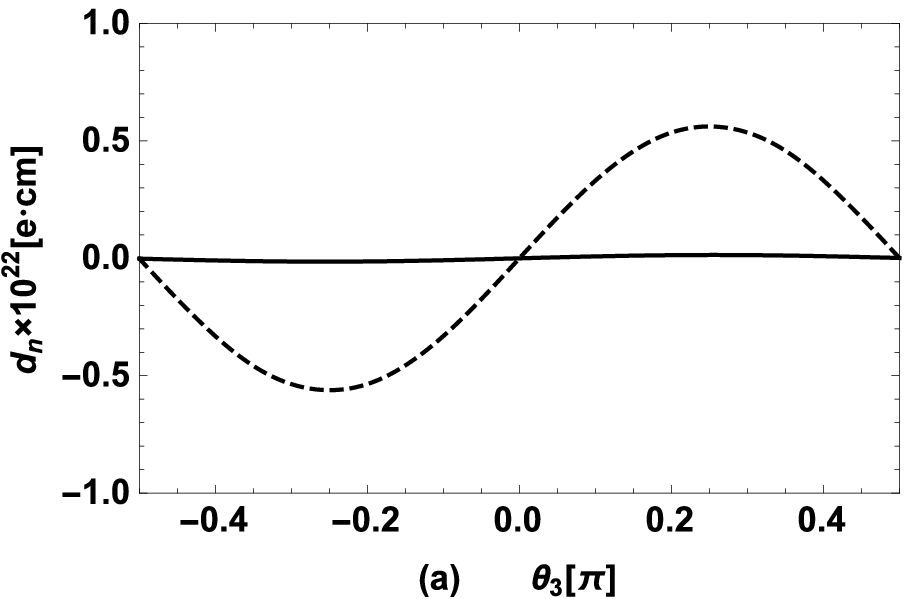}%
\vspace{0.5cm}
\includegraphics[width=2.7in]{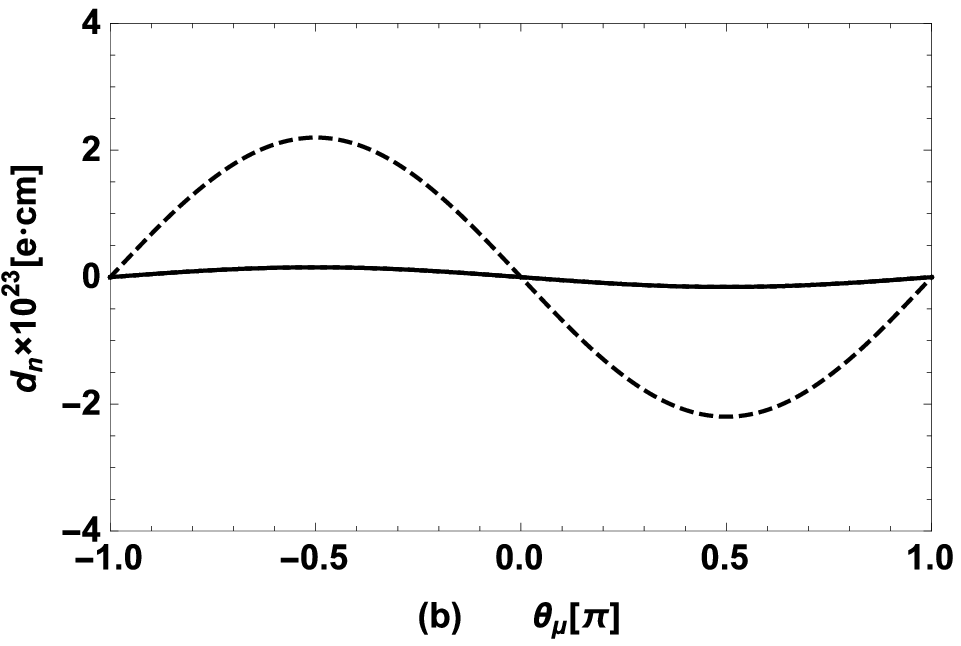}
\vspace{0cm}
\par
\hspace{-0.in}
\includegraphics[width=2.7in]{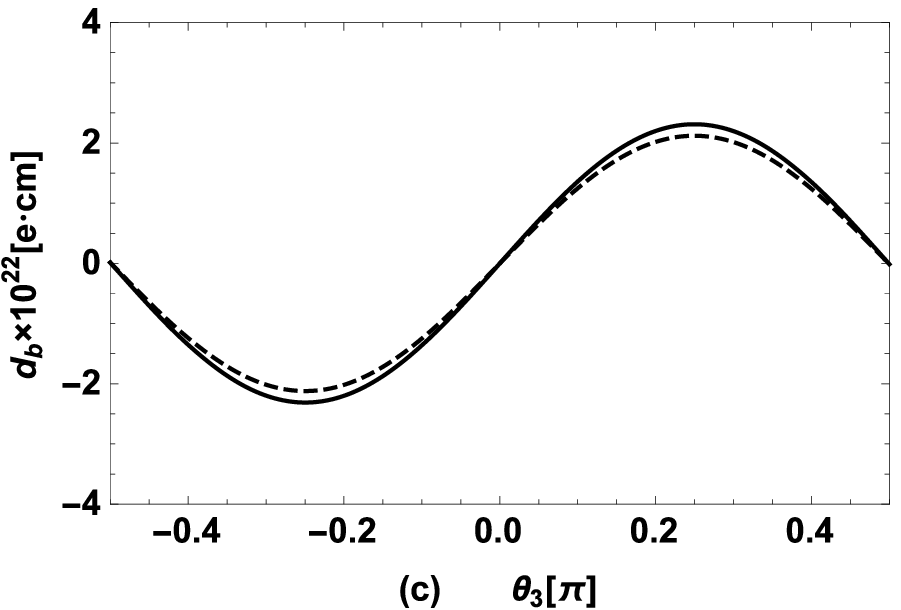}%
\vspace{0.5cm}
\includegraphics[width=2.7in]{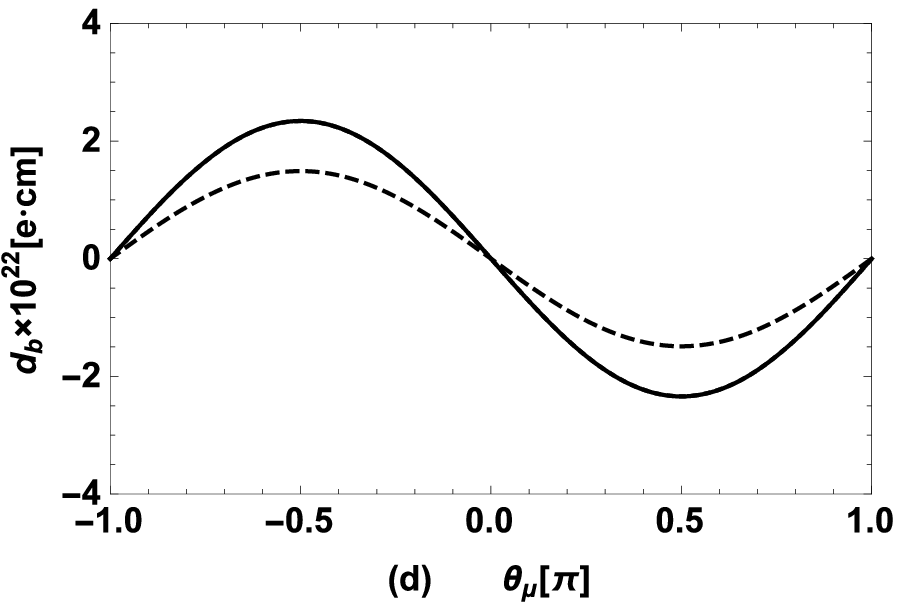}
\vspace{0cm}
\par
\hspace{-0.in}
\includegraphics[width=2.7in]{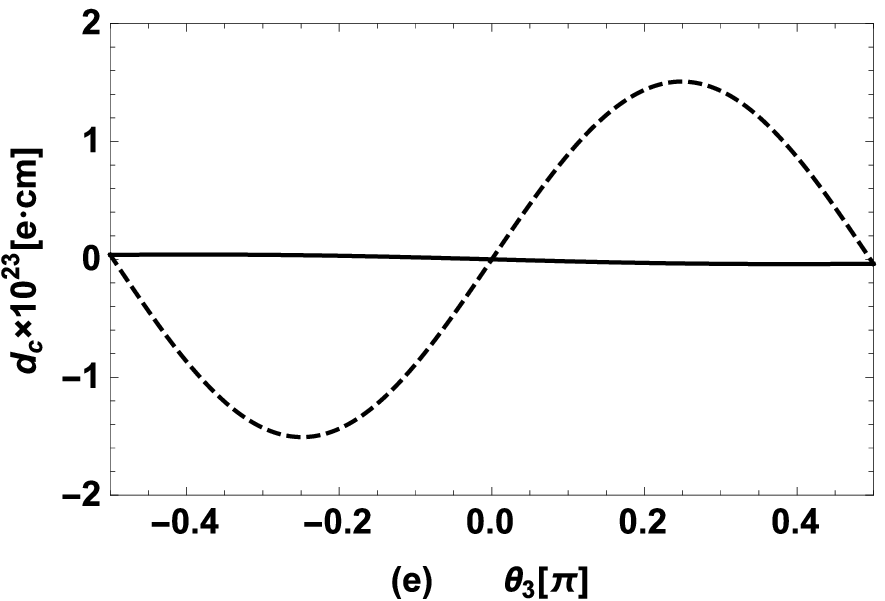}%
\vspace{0.5cm}
\includegraphics[width=2.7in]{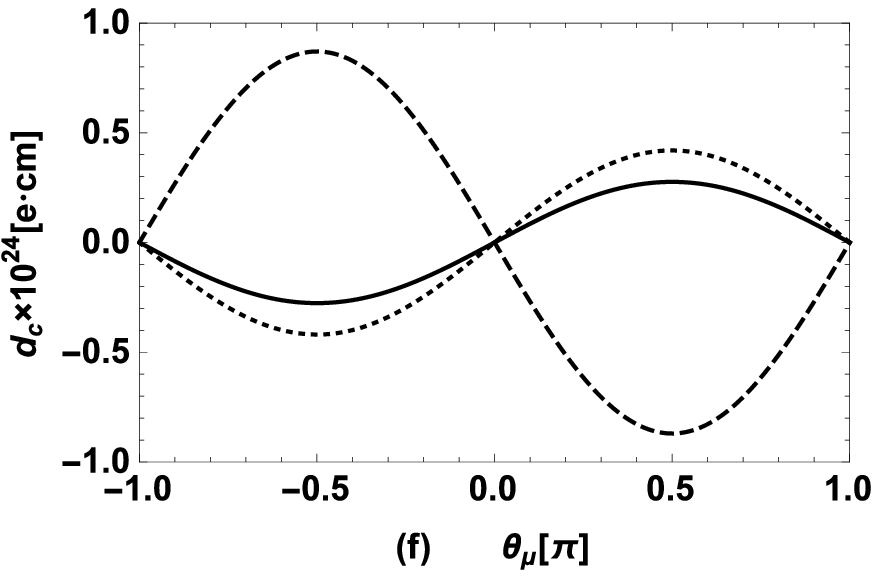}
\vspace{0cm}
\caption[]{$d_n$, $d_b$, $d_c$ versus $\theta_3$, $\theta_\mu$ are plotted, where the solid, dashed and dotted lines denote the one loop results, the sum of one-loop and two-loop gluino results, the sum of one-loop and two-loop Barr-Zee type results respectively.}
\label{theta3mu}
\end{figure}

The two-loop gluino corrections are included in our calculation, $\theta_3$ (the phase of gluino mass $M_3$) can make contributions to $d_n$, $d_b$ and $d_c$ through these two-loop gluino diagrams. In addition, the $\mu$ term makes the dominant contributions to the EWB, and the corresponding CPV phase $\theta_\mu$ is requested to be large. Meanwhile, $\theta_\mu$ can make contributions to these EDMs through both the two-loop gluino and Barr-Zee type diagrams. Hence, the effects of $\theta_3$ and $\theta_\mu$ to these EDMs are interesting. In order to explore the effects of $\theta_3$, $\theta_\mu$ and the relative corrections of two-loop diagrams to the one-loop results, we take $M_1=\frac{1}{2}M_2=\frac{1}{2}M_{B'}=\frac{1}{2}M_{BB'}=\frac{1}{2}\mu=0.3{\;\rm TeV}$, $M_3=0.3{\;\rm TeV}$, $\mu'=0.8{\;\rm TeV}$, $\tan\beta=10$, $\tan\beta'=1.15$, $g_{_B}=0.4$, $g_{_{YB}}=-0.4$, $M_Q=2$, $A_Q=0.1$ and all other CPV phases to be zero. In the following analysis, these parameters are fixed at the above values for simplicity when we do not consider them as variables to explore their effects. The parameter space we chosen above can satisfy the constraints from experiments~\cite{PDG}. Then we plot $d_n$ versus $\theta_3$, $\theta_\mu$ in Fig.~\ref{theta3mu}(a), (b) respectively, where the solid, dashed and dotted lines denote the one-loop result, the sum of one-loop and two-loop gluino results, the sum of one-loop and two-loop Barr-Zee type results respectively. There is no dotted line in Fig.~\ref{theta3mu}(a), because $\theta_3$ do not make contributions through Barr-Zee type diagrams. Similarly, $d_b$ versus $\theta_3$, $\theta_\mu$ are plotted in Fig.~\ref{theta3mu}(c), (d), and $d_c$ versus $\theta_3$, $\theta_\mu$ are plotted in Fig.~\ref{theta3mu}(e), (f).

Fig.~\ref{theta3mu}(c) shows that the relative corrections of two-loop gluino diagrams to the one-loop result of $d_b$ can reach $\sim10\%$ when the CPV contributions only come from $\theta_3$, which produces more precise prediction of $d_b$. From Fig.~\ref{theta3mu}(a), (e) we can see that the two-loop gluino corrections make dominant contributions to $d_{n,c}$ with respect to the one-loop results. Combining with the results shown in Fig.~\ref{theta3mu}(b), (d), (f), we can conclude that the two-loop gluino corrections can make more important contributions to lighter quark, and the two-loop gluino corrections to $d_n$, $d_c$ are even larger than the corresponding one-loop results. When the CPV contributions only come from $\theta_3$ or $\theta_\mu$, the dominant contributions of one-loop and two-loop gluino corrections to quark EDMs come from charginos, and Yukawa coupling constant appears in the right-hand part of the interaction vertex chargino-quark-squark. Hence, the expression of $d_q^{\gamma,g(3)}$ in Eq.~(\ref{17}) shows that the contributions from charginos are proportional to Yukawa coupling constant. However, for the two-loop gluino corrections when charginos appear as internal particles in Eq.~(\ref{21}), there is a term does not depend on Yukawa coupling constant, this is why the two-loop gluino diagrams can make dominant corrections to $d_n$ or $d_c$. In addition, Fig.~\ref{theta3mu}(b), (d) show that the corrections of Barr-Zee type diagrams to $d_{n,b}$ are negligible, while the relative corrections of Barr-Zee type diagrams to the one-loop result of $d_{c}$ can reach $50\%$. It also can be noted that the most strict constraints on $\theta_3$ and $\theta_\mu$ come from the experimental upper bound on $d_n$. And the contributions to $d_n$ from $\theta_3$ is larger than the contributions from $\theta_\mu$, hence the contributions from large $\theta_\mu$ which is needed to generate the baryon asymmetry can be cancelled by appropriate $\theta_3$. The contributions to $d_n$ from $\theta_\mu$ can be cancelled by the one of $\theta_3$ results from the fact that $\theta_\mu$ affects the numerical results mainly through the squark mass matrixes, but the effects of it are suppressed by the corresponding Yukawa coupling constants which can be seen in Eq.~(\ref{A1}, \ref{A3}). On the other hand, $\theta_3$ appears in the analysis result directly and there is no suppressive factor. It can also explain that the contributions to $d_b$ from $\theta_\mu$ are comparable with the ones from $\theta_3$ while the contributions to $d_{n,c}$ from $\theta_\mu$ are smaller than the ones from $\theta_3$, which can be seen directly by comparing Fig.~\ref{theta3mu}(a), (c), (e) with Fig.~\ref{theta3mu}(b), (d), (f). Finally, we can also note that the theoretical prediction of $d_{c}$ is well below the present upper bound in our chosen parameter space which indicates the present upper bound on $d_{c}$ limits the parameter space of the B-LSSM weakly.

\begin{figure}
\setlength{\unitlength}{1mm}
\centering
\includegraphics[width=2.7in]{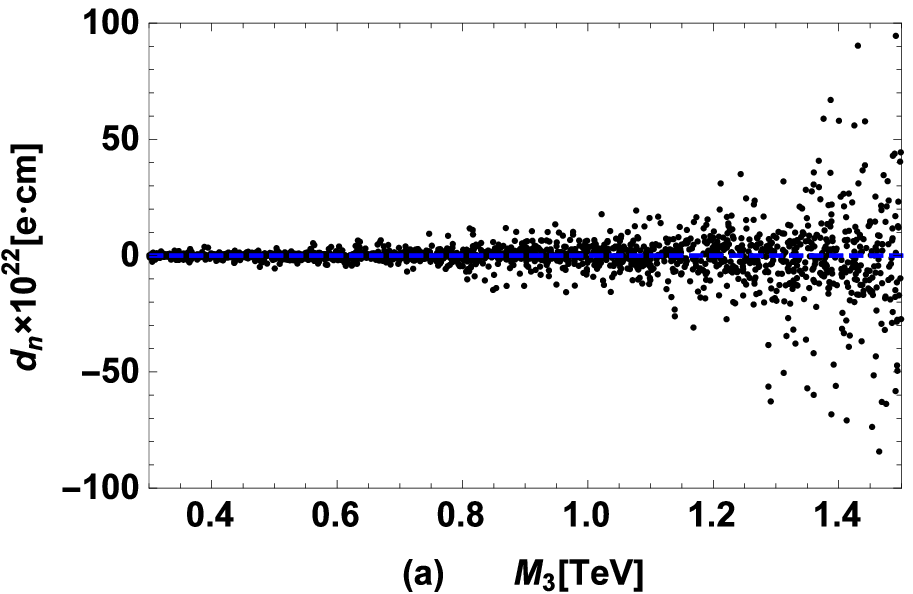}%
\vspace{0.5cm}
\includegraphics[width=2.7in]{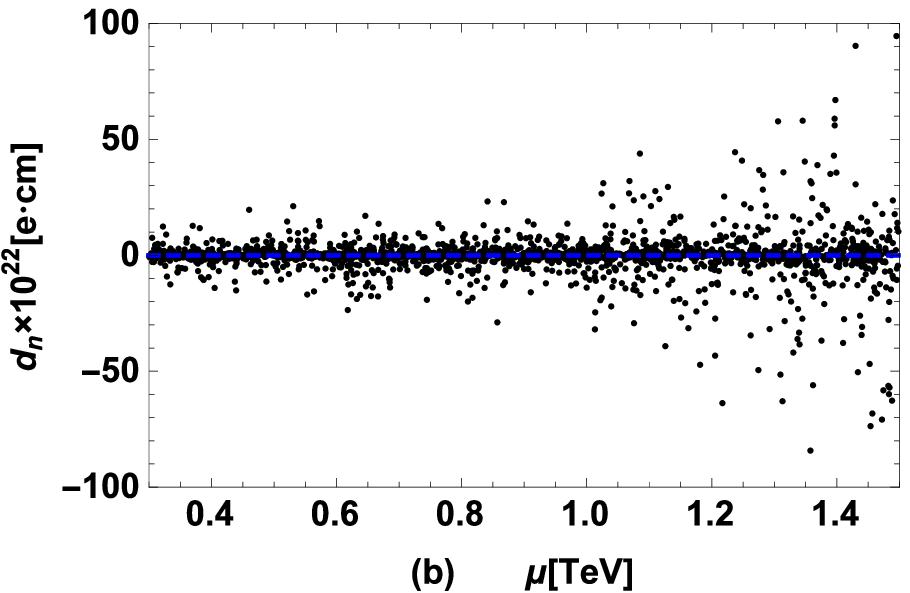}
\vspace{0cm}
\par
\hspace{-0.in}
\includegraphics[width=2.7in]{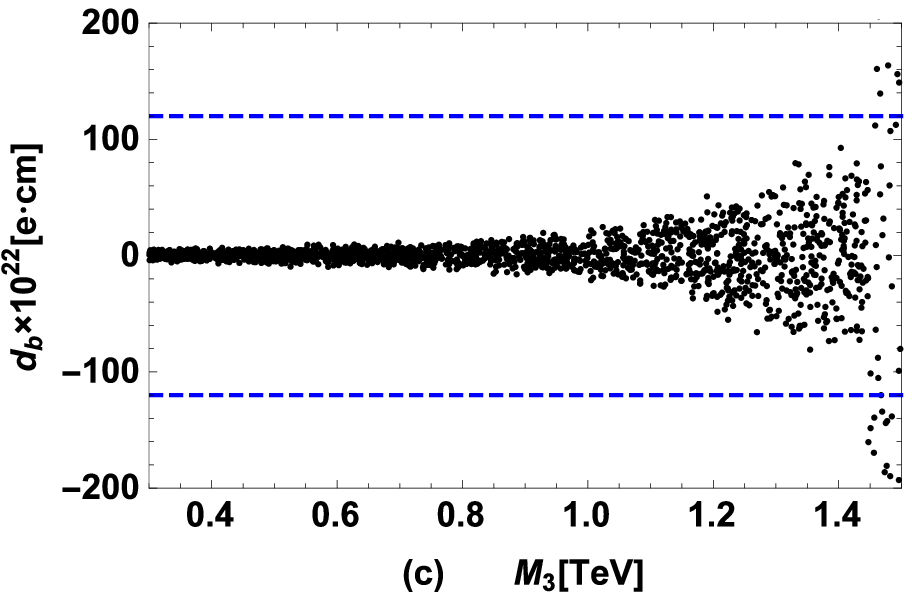}%
\vspace{0.5cm}
\includegraphics[width=2.7in]{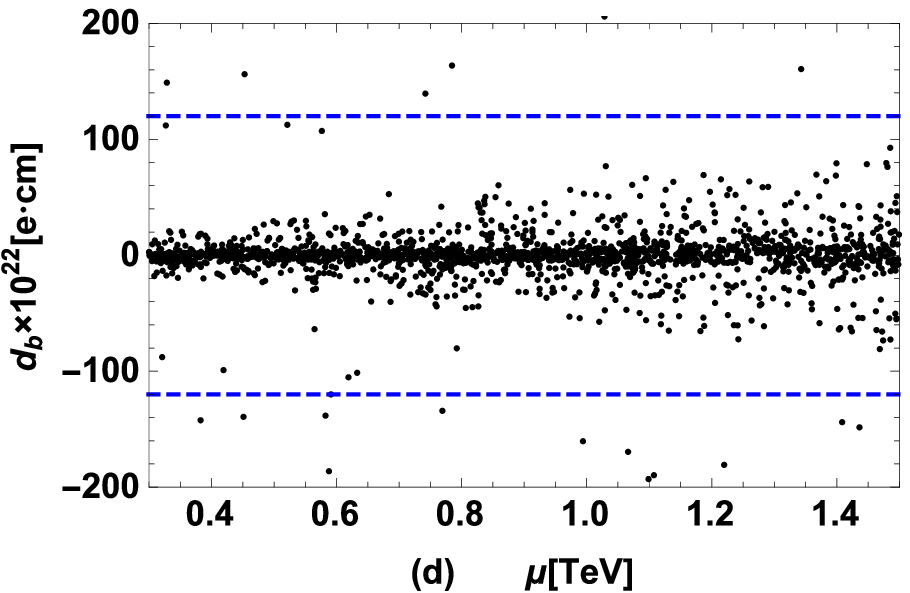}
\vspace{0cm}
\par
\hspace{-0.in}
\includegraphics[width=2.7in]{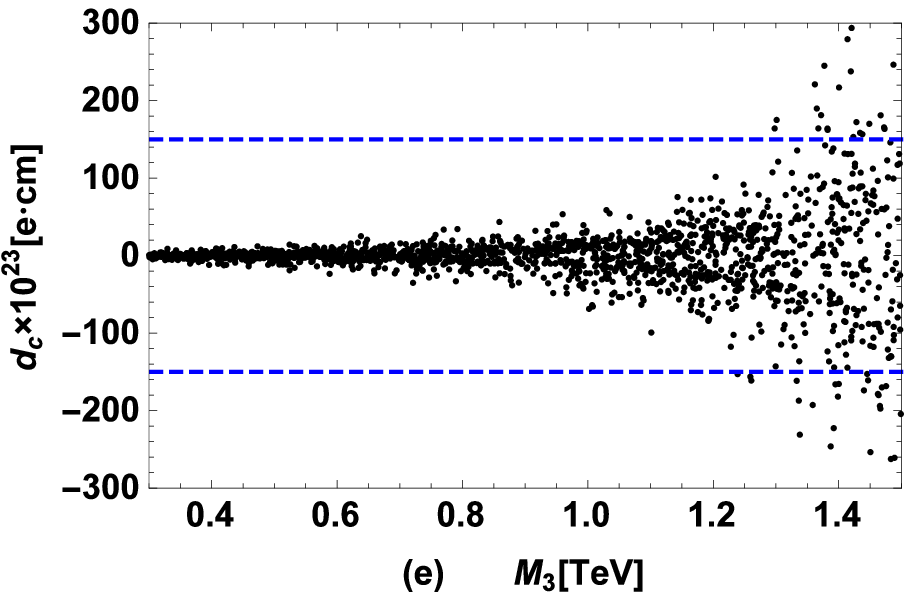}%
\vspace{0.5cm}
\includegraphics[width=2.7in]{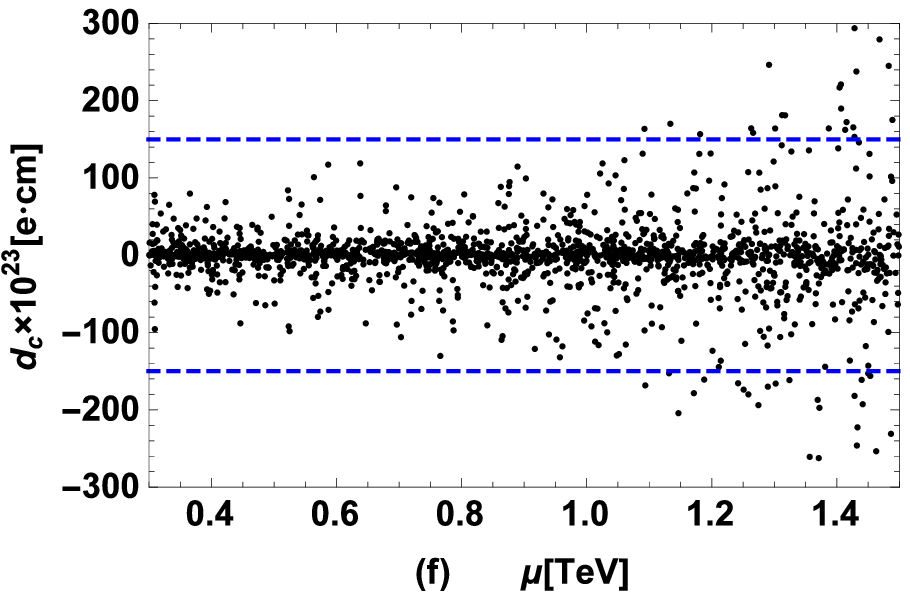}
\vspace{0cm}
\caption[]{Scanning the parameter space shown in Eq.~(\ref{R4}), $d_n$, $d_b$ and $d_c$ versus $M_3$, $\mu$ are plotted, where the blue dashed lines denote the corresponding upper bounds.}
\label{Scancel}
\end{figure}
It can be noted from Fig.~\ref{theta3mu} that the CPV contributions from $M_3$ to these EDMs can cancel the ones from $\mu$ which is requested to be large for the taking place of EWB. However, $M_3$ and $\mu$ are fixed because we focus on the relative corrections of gluino and Barr-Zee type two-loop diagrams in Fig.~\ref{theta3mu}. In order to explore the effects of $M_3$, $\mu$ and the corresponding CPV phases clearly, we scan the following parameter space:
\begin{eqnarray}
&&M_{3}=(0.3,\;1.5),\;\mu=(0.3,\;1.5),\;\theta_{3}=(-0.5\pi,\;0.5\pi),\;\theta_{\mu}=(-\pi,\;\pi).\label{R4}
\end{eqnarray}
We take $M_{3},\;\mu\gtrsim0.3$ TeV in order to avoid the ranges excluded by the experiments~\cite{PDG}. Since both the two-loop gluino and Barr-Zee type diagrams can make important contributions to the numerical results, we take the summing of all contributions from these two-loop and one-loop corrections in the following analysis. Then $d_n$ versus $M_3$, $\mu$ are plotted in Fig.~\ref{Scancel}(a), (b) respectively, where the blue dashed lines denote the experimental upper bound on $d_n$. Similarly, $d_b$, $d_c$ versus $M_3$, $\mu$ are plotted in Fig.~\ref{Scancel}(c, d), (e, f). Fig.~\ref{Scancel}(a, c, e) show that $\mu$, $\theta_3$, $\theta_\mu$ affect the numerical results more obviously with large $M_3$. Especially when $M_3\gtrsim1$ TeV, the theoretical predictions of these EDMs can be larger than the corresponding results shown in Fig.~\ref{theta3mu} by about two orders of magnitude. Comparing with the results shown in Fig.~\ref{Scancel}(b, d, f), these parameters affect the numerical results more obviously when $\mu$ is large, but the trend is less obvious than the results shown in Fig.~\ref{Scancel}(a, c, e). It indicates that $M_3$ affect the numerical results more obvious than $\mu$, and the contributions to $d_n$ from $\theta_\mu$ can be cancelled by appropriate $\theta_3$ which is shown in Fig.~\ref{Scancel}(a, b). In addition, the experimental upper bounds on $d_b$, $d_c$ also limit the parameter space when $M_3$, $\mu$ are large as we can note in Fig.~\ref{Scancel}(c, d, e, f).

\begin{figure}
\setlength{\unitlength}{1mm}
\centering
\includegraphics[width=2.7in]{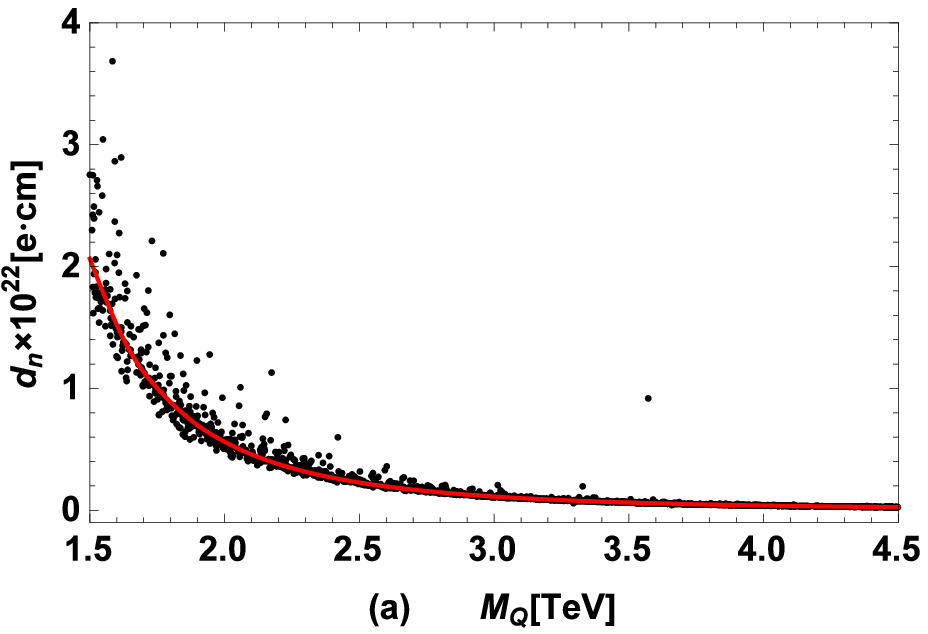}%
\vspace{0.5cm}
\includegraphics[width=2.7in]{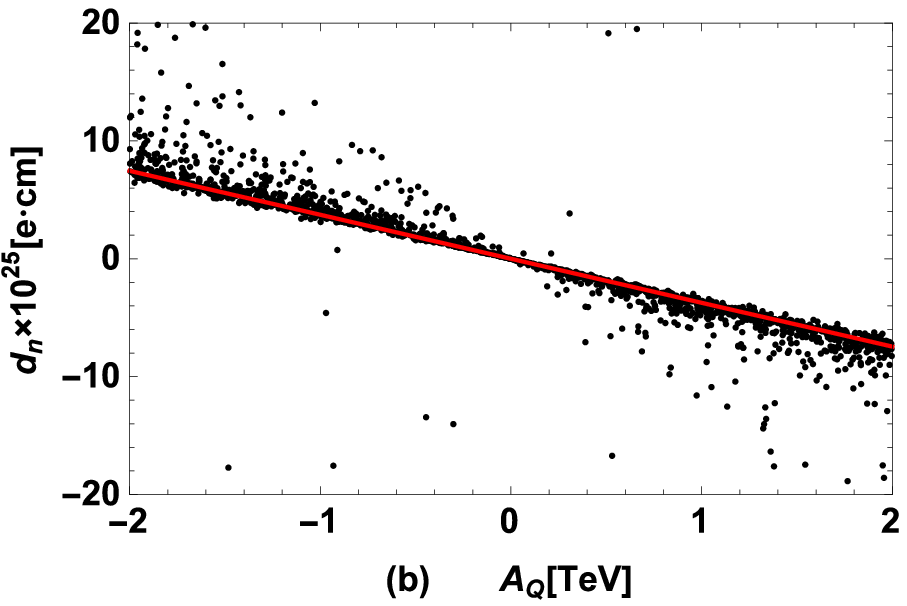}
\vspace{0cm}
\par
\hspace{-0.in}
\includegraphics[width=2.7in]{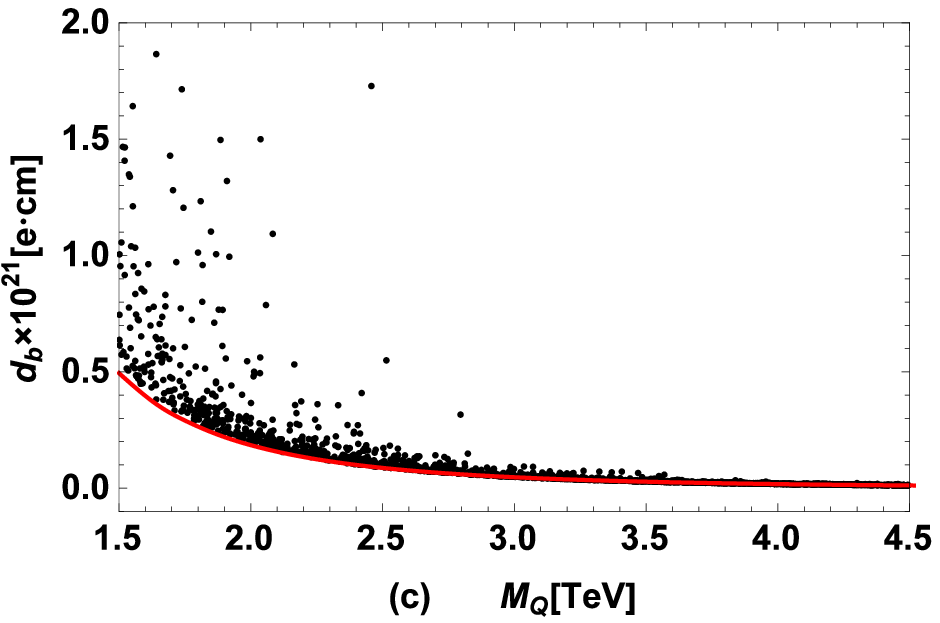}%
\vspace{0.5cm}
\includegraphics[width=2.7in]{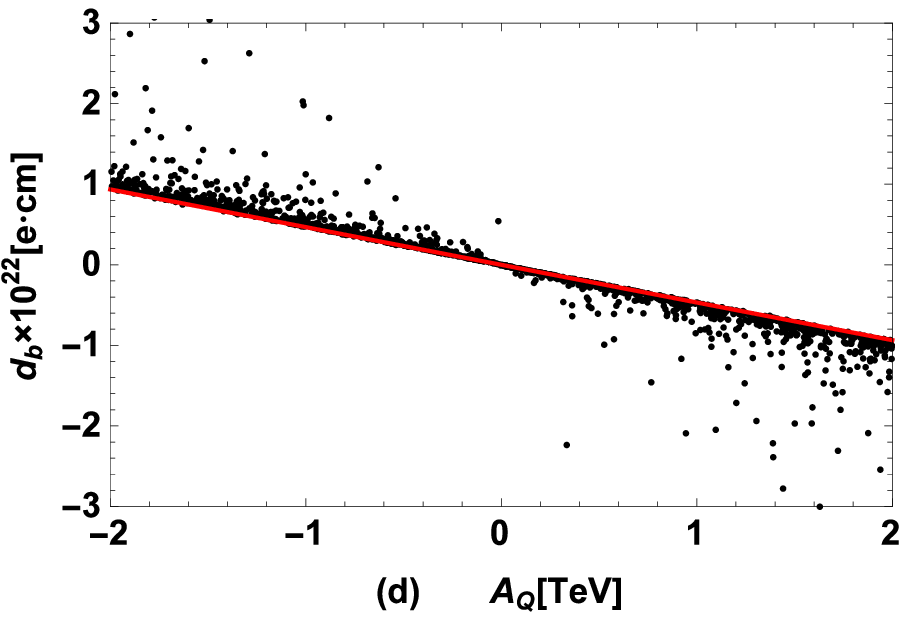}
\vspace{0cm}
\par
\hspace{-0.in}
\includegraphics[width=2.7in]{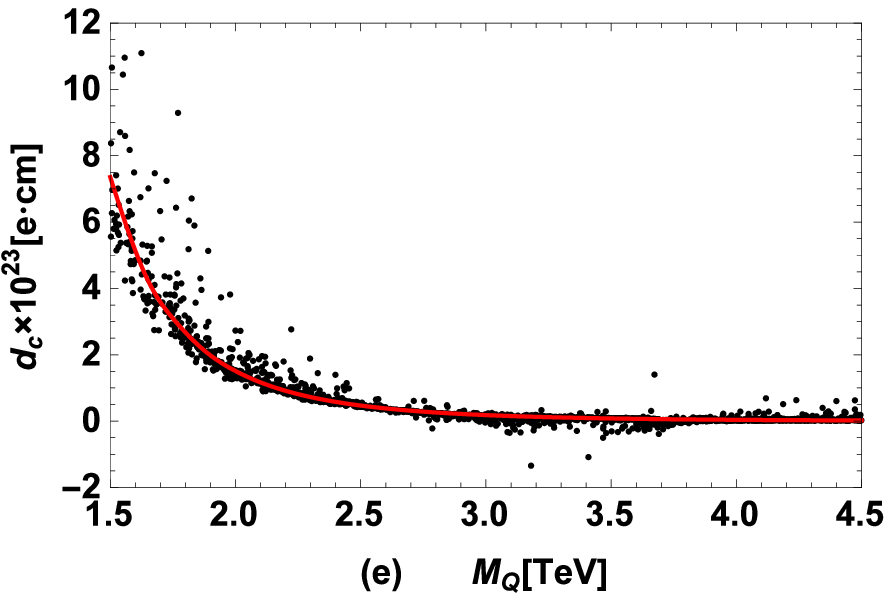}%
\vspace{0.5cm}
\includegraphics[width=2.7in]{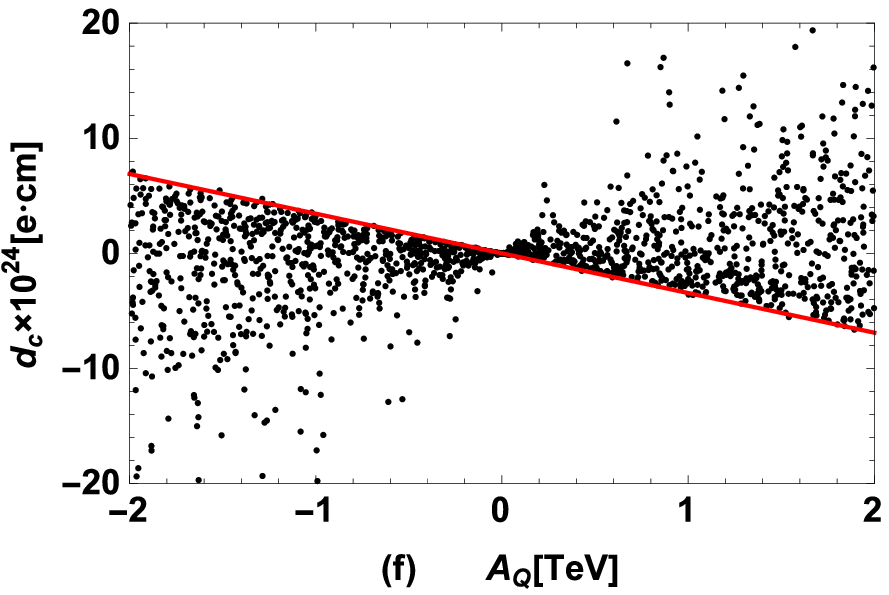}
\vspace{0cm}
\caption[]{Scanning the parameter space shown in Eq.~(\ref{R1}), (\ref{R2}), $d_n$ versus $M_Q$, $A_Q$ are plotted respectively, where the red lines denote the corresponding MSSM predictions in the same parameter space. Similarly, $d_{b,c}$ versus $M_Q$, $A_Q$ are also plotted.}
\label{M}
\end{figure}
With respect to the MSSM, there are three new parameters $\tan\beta'$, $g_{_B}$ and $g_{_{YB}}$ in the B-LSSM, which can also affect the theoretical predictions of $d_{n,b,c}$. Taking $\theta_3=0.25\pi$, we scan the parameter space below to explore the effects of $M_Q$, $\tan\beta'$, $g_{_B}$ and $g_{_{YB}}$:
\begin{eqnarray}
&&M_Q=(1.5,\;4.5),\;\tan\beta'=(1.1,\;1.5),\;g_{_B}=(0.1,\;0.7),\;g_{_{YB}}=(-0.7,\;-0.1).\label{R1}
\end{eqnarray}
We choose $\tan\beta'\geq1.1$ because the region $\tan\beta'<1.1$ is excluded by the SM-like Higgs boson mass easily~\cite{JLYang:2018}. And the B-LSSM decouples to the MSSM when $g_{_B}$, $|g_{_{YB}}|$ are too small, hence we take $g_{_B}>0.1$, $g_{_{YB}}<-0.1$ respectively. Then $d_n$ versus $M_Q$ are plotted in Fig.~\ref{M}(a), the MSSM prediction of $d_n$ in the same parameter space is also plotted (red line) in order to compare with the MSSM predictions clearly. Similarly, $d_b$ and $d_c$ versus $M_Q$ are plotted in Fig.~\ref{M}(c), (e). In addition, the trilinear scalar terms $T_{u,d}$ can also have CPV phase $\theta_A$. Assuming $M_Q=2$, $\theta_A=0.5\pi$ and all other phases are zero, we scan the following parameter space:
\begin{eqnarray}
&&A_Q=(-2,\;2),\;\tan\beta'=(1.1,\;1.5),\;g_{_B}=(0.1,\;0.7),\;g_{_{YB}}=(-0.7,\;-0.1).\label{R2}
\end{eqnarray}
We take $|A_Q|<2$ TeV approximately in order to satisfy the Color or Charge Breaking bounds~\cite{Gunion:1987qv}. Then $d_n$ versus $A_Q$ are plotted in Fig.~\ref{M}(b), the MSSM prediction of $d_n$ in the same parameter space is plotted as red line. Similarly, $d_b$ and $d_c$ versus $A_Q$ are plotted in Fig.~\ref{M}(d), (f).

Fig.~\ref{M}(a), (c), (e) show that the contributions to $d_{n,b,c}$ decrease with the increasing of $M_Q$ which coincides with the decoupling theorem. With the increasing of squark mass parameter $M_Q$, new parameters $\tan\beta'$, $g_{_B}$, $g_{_{YB}}$ affect the numerical results more weakly. Because these new parameters affect the results mainly through affecting the squark masses, the effects of them are suppressed by large $M_Q$. It can be noted that $\tan\beta'$, $g_{_B}$, $g_{_{YB}}$ can affect the numerical results obviously when $M_Q\lesssim2.5$ TeV. And for the same choice of $M_Q$, the theoretical predictions of these EDMs in the B-LSSM can be about twice larger than the ones in the MSSM for appropriate $\tan\beta'$, $g_{_B}$, $g_{_{YB}}$. For the points that $\tan\beta'$, $g_{_B}$, $g_{_{YB}}$ make the squark masses heavier, the predictions of these EDMs are smaller than the ones in the MSSM according to the decoupling theorem. As we can see from Fig.~\ref{M}(a), (e), there are some points below the MSSM predictions. However, there are no points below the MSSM predictions for $d_b$. Because the Yukawa coupling constant for $b$ quark is large, and new introduced neutralinos in the B-LSSM can make large contributions to $d_b$ through the mixing with Higgsino, which can compensate the suppressive role of heavier squark.

From Fig.~\ref{M}(b), (d), (f) we can see that $\theta_A$ can also make important contributions to these EDMs, and the signs of $d_{n,b,c}$ can be changed when we take the sign of $A_Q$ be opposite. In addition, Fig.~\ref{M}(b), (d) show that the predictions of $d_{n,b}$ in B-LSSM are larger than the ones in MSSM for most points, while the prediction of $d_{c}$ in B-LSSM is smaller than the ones in MSSM for most points and the sign of $d_c$ are changed for some points. The dominant contributions to $d_q$ come from $d_q^\gamma$ when $M_Q=2$, and $d_q^\gamma$ increases more slowly than $d_q^g$ with the decreasing of squark masses. For $c$ quark, the signs of $d_c^\gamma$ and $d_c^g$ are opposite, and the contributions from $d_q^\gamma$ are larger than $d_q^g$ when the squark masses are heavy. As a result, all points in Fig.~\ref{M}(e) below the corresponding MSSM prediction, the sign of $d_c$ can be changed and $|d_c|$ can even be larger than the corresponding MSSM prediction.

\begin{figure}
\setlength{\unitlength}{1mm}
\centering
\includegraphics[width=2.7in]{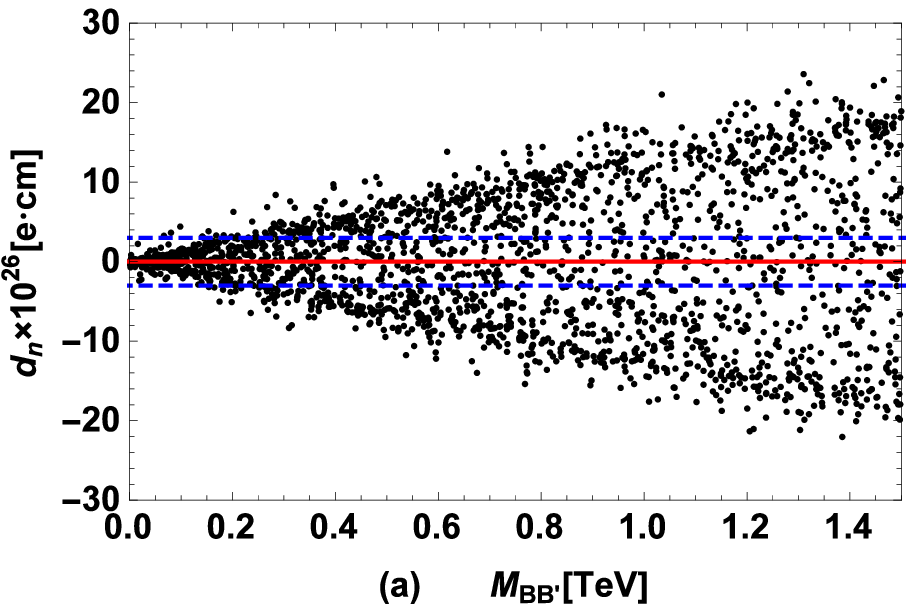}%
\vspace{0.5cm}
\includegraphics[width=2.7in]{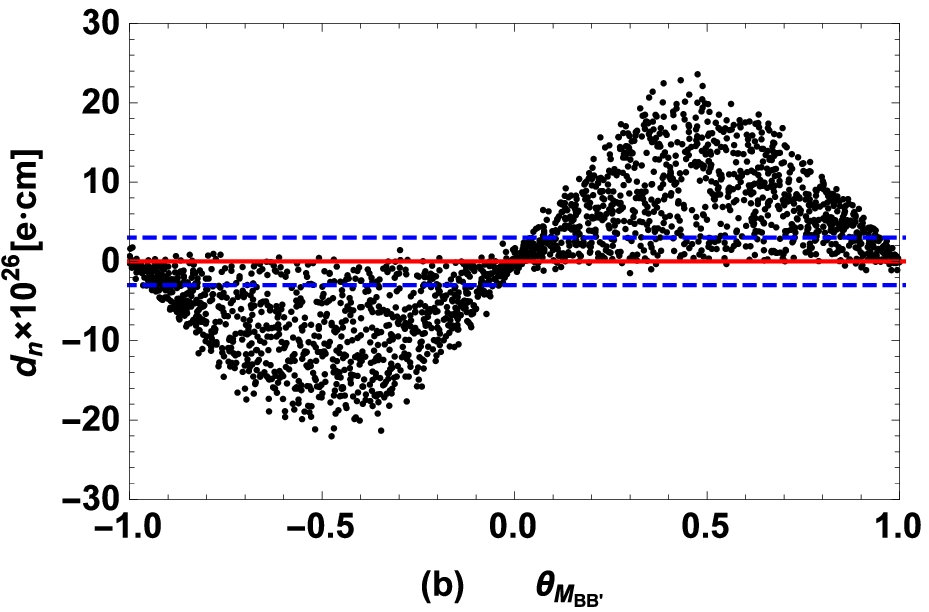}
\vspace{0cm}
\par
\hspace{-0.in}
\includegraphics[width=2.7in]{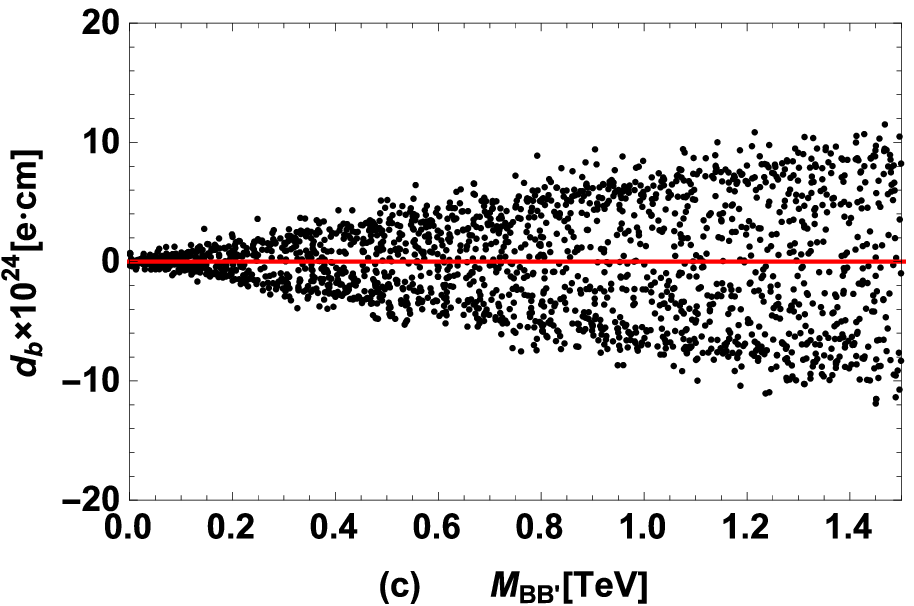}%
\vspace{0.5cm}
\includegraphics[width=2.7in]{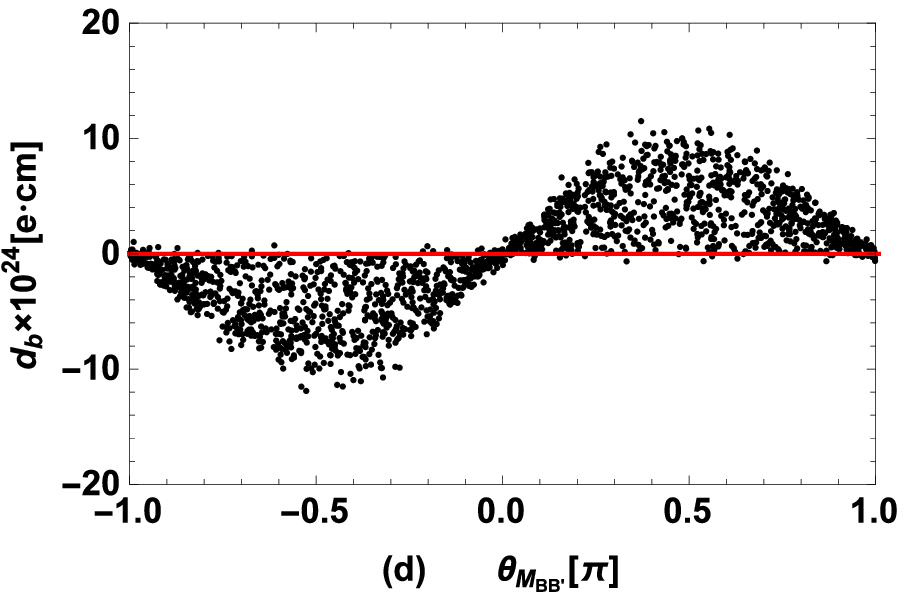}
\vspace{0cm}
\par
\hspace{-0.in}
\includegraphics[width=2.7in]{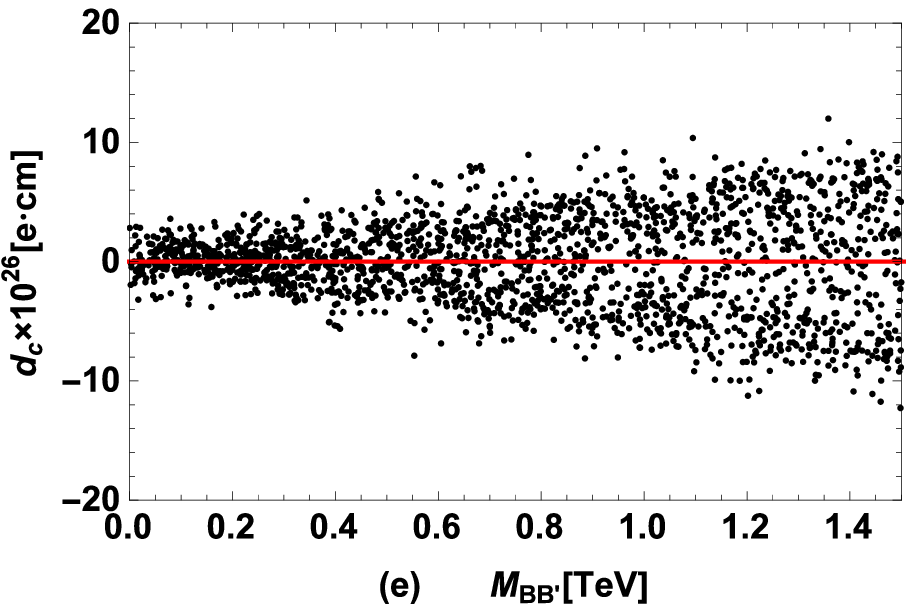}%
\vspace{0.5cm}
\includegraphics[width=2.7in]{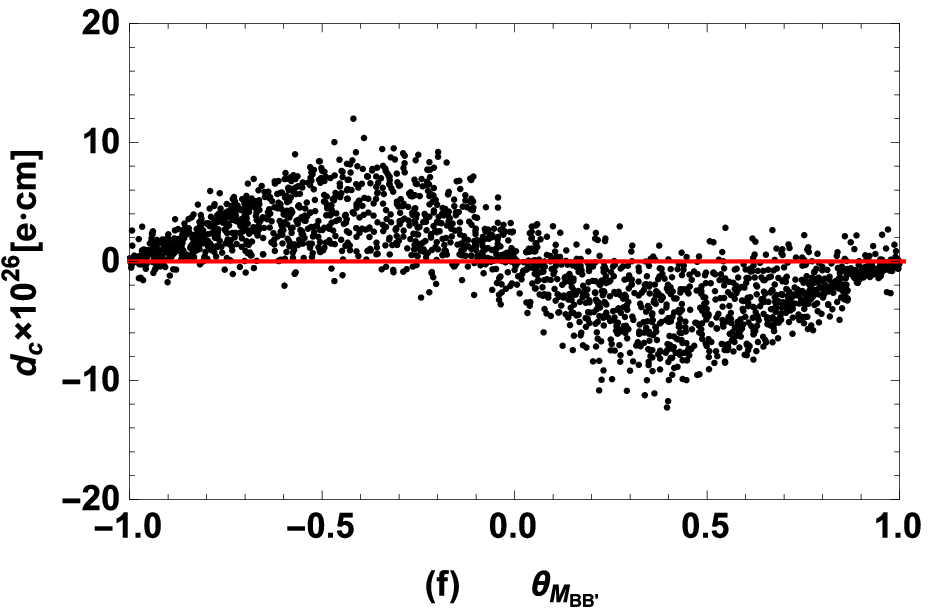}
\vspace{0cm}
\caption[]{Scanning the parameter space shown in Eq.~(\ref{R3}), $d_n$, $d_b$, $d_c$ versus $M_{BB'}$, $\theta_{M_{BB'}}$ are plotted, where the red lines denote the corresponding MSSM predictions in the same parameter space, the blue dashed lines denote the experimental upper bound on $d_n$.}
\label{thetaBL}
\end{figure}
Comparing with the MSSM, there are two new mass terms $M_{BB'}$ and $M_{B'}$ in the B-LSSM, which can also have CPV phases $\theta_{M_{BB'}}$, $\theta_{M_{B'}}$ respectively. Both of $M_{BB'}$ and $M_{B'}$ can be very small and the gaugino masses are still large enough to satisfy the experimental lower bounds on gaugino masses. Assuming all contributions from other phases are cancelled each other completely, and the contributions to these EDMs come from $\theta_{M_{BB'}}$, $\theta_{M_{B'}}$. We scan the parameter space below to explore the effects of $M_{BB'}$, $M_{B'}$ and corresponding CPV phases $\theta_{M_{BB'}}$, $\theta_{M_{B'}}$:
\begin{eqnarray}
&&M_{BB'}=(0,\;1.5),\;M_{B'}=(0,\;1.5),\;\theta_{M_{BB'}}=(-\pi,\;\pi),\;\theta_{M_{B'}}=(-\pi,\;\pi).\label{R3}
\end{eqnarray}
We take $M_{BB'},\;M_{B'}\lesssim1.5$ TeV because the results are limited strictly by the present upper bound on $d_n$ when $M_{BB'}$, $M_{B'}$ are too large. Then $d_n$ versus $M_{BB'}$, $\theta_{M_{BB'}}$ are plotted in Fig.~\ref{thetaBL}(a), (b) respectively, where the blue dashed lines denote the experimental upper bound on $d_n$, and the red lines denote the corresponding MSSM predictions in the same parameter space. The MSSM prediction of $d_n$ is $0$ in this case because there are no such CPV parameters in the MSSM. Similarly, we plot $d_b$, $d_c$ versus $M_{BB'}$, $\theta_{M_{BB'}}$ in Fig.~\ref{thetaBL}(c, d), (e, f) respectively. There are no blue dashed lines in Fig.~\ref{thetaBL}(c, d), (e, f) because the upper bounds on $d_b$, $d_c$ are much larger than the results shown in the picture.

Fig.~\ref{thetaBL}(a, b) show that the theoretical prediction of $d_n$ in the B-LSSM can also exceed the present upper bound on $d_n$ when the CPV contributions come from new mass parameters $M_{BB'}$, $M_{B'}$. When $\theta_{M_{BB'}}\gtrsim0.05\pi$, the range of $M_{BB'}$ is limited by the present upper bound on $d_n$. And the upper bound on $d_n$ does not limit the ranges of $\theta_{M_{BB'}}$, $M_{B'}$, $\theta_{M_{B'}}$ when $M_{BB'}\lesssim0.2$ TeV. In addition, the theoretical predictions of $d_b$, $d_c$ can reach $10^{23}$, $10^{25}$ respectively when the contributions come from new CPV effects in the B-LSSM which may be detected in the near future. It can also be noted that the effects of $M_{B'}$, $\theta_{M_{B'}}$ are suppressed by small $M_{BB'}$ or $\theta_{M_{BB'}}$. $M_{BB'}$ is the mixing term of $\tilde\lambda_B$, $\tilde\lambda_{B'}$, hence $M_{B'}$ can make large contributions to these EDMs through the channel of $\tilde\lambda_{B'}$ and the mixing with $\tilde\lambda_B$ when $M_{BB'}$, $\theta_{M_{BB'}}$ are large. As $M_{BB'}$ and $\theta_{M_{BB'}}$ approach to zero, $M_{B'}$ contributes to these EDMs only through the channel of $\tilde\lambda_{B'}$ which is suppressed by the heavy neutralino mass correspondingly. As a result, $M_{BB'}$ and $\theta_{M_{BB'}}$ affect the numerical results more obviously than $M_{B'}$ and $\theta_{M_{B'}}$.

\section{Summary\label{sec5}}

In this work, we explore the EDMs of neutron $d_n$ and heavy quarks $d_b$, $d_c$ in the B-LSSM. Some two loop gluino and Barr-Zee type diagrams are included in the calculation. Considering the constraints from updated experimental data, the numerical results show that the two-loop gluino diagrams can make important corrections to $d_{n,b,c}$, while the two-loop Barr-Zee type diagrams can make important corrections to $d_{c}$. The most strict constraints on $\theta_3$ and $\theta_\mu$ come from the experimental upper bound on $d_n$, and the contributions from large $\theta_\mu$ which is needed for the taking place of EWB can be cancelled by the contributions from $\theta_3$. With respect to the MSSM, there are new parameters $\tan\beta'$, $g_{_B}$, $g_{_{YB}}$ in the B-LSSM can affect the numerical results by affecting the squark masses and the mixing in the neutralino sector. When the squark mass parameter $M_Q<2.5$ TeV, these new parameters affect the numerical results obviously and increase the theoretical predictions of these EDMs in the MSSM by about twice in the same parameter space. In addition, there are new CPV phases of two additional mass terms $M_{B'}$, $M_{BB'}$ in the neutralino sector. These new CPV phases are absent in the MSSM and can make contributions to these EDMs through both one-loop and two-loop corrections. When these new CPV sources in the B-LSSM are considered, the numerical results show that the theoretical predictions of $d_b$, $d_c$ can reach $10^{23}$, $10^{25}$ respectively, which may be detected in the near future.

\begin{acknowledgments}

The work has been supported by the National Natural Science Foundation of China (NNSFC) with Grants No. 11535002, No. 11647120, and No. 11705045; Peng-Huan-Wu Theoretical Physics Innovation Center: 11947302; Natural Science Foundation of Hebei province with Grants No. A2016201010 and No. A2016201069; the youth top-notch talent support program of the Hebei Province; Hebei Key Lab of Optic-Eletronic Information and Materials; the Midwest Universities Comprehensive Strength Promotion project.
\end{acknowledgments}

\appendix

\section{New definitions of some mass matrixes in the B-LSSM. \label{mass matrix}}
On the basis $(\tilde d_L, \tilde d_R)$, new definition of the mass matrix for down type squarks is given by
\begin{eqnarray}
&&m_{\tilde d}^2=
\left(\begin{array}{cc}m_{dL},&\frac{1}{\sqrt2}(v_1 T_d^\dagger-v_2\mu Y_d^\dagger)\\\frac{1}{\sqrt2}(v_1 T_d-v_2\mu^* Y_d),&m_{dR}\end{array}\right),\label{A1}
\end{eqnarray}
\begin{eqnarray}
&&m_{dL}=\frac{1}{24}\Big(2g_{_B}(g_{_B}+g_{_{YB}})(u_2^2-u_1^2)+(3g_2^2+g_1^2+g_{_{YB}}^2+
g_{_B}g_{_{YB}})(v_2^2-v_1^2)\Big)\nonumber\\
&&\qquad\;\quad\;+m_{\tilde q}^2+\frac{v_1^2}{2}Y_d^\dagger Y_d,\nonumber\\
&&m_{dR}=\frac{1}{24}\Big(2g_{_B}(g_{_B}-2g_{_{YB}})(u_1^2-u_2^2)+2(g_1^2+g_{_{YB}}^2-
\frac{1}{2}g_{_B}g_{_{YB}})(v_2^2-v_1^2)\Big)\nonumber\\
&&\qquad\;\quad\;+m_{\tilde d}^2+\frac{v_1^2}{2}Y_d^\dagger Y_d,\label{massd}
\end{eqnarray}
which can be diagonalized by $Z^{\tilde d}$.

On the basis $(\tilde u_L, \tilde u_R)$, new definition of the mass matrix for up type squarks is given by
\begin{eqnarray}
&&m_{\tilde u}^2=
\left(\begin{array}{cc}m_{uL},&\frac{1}{\sqrt2}(v_2 T_u^\dagger-v_1\mu Y_u^\dagger)\\\frac{1}{\sqrt2}(v_2 T_u-v_1\mu^* Y_u),&m_{uR}\end{array}\right),\label{A3}
\end{eqnarray}
where,
\begin{eqnarray}
&&m_{uL}=m_{\tilde q}^2+\frac{1}{24}\Big[2g_{_B}(g_{_B}+g_{_{YB}})(u _2^2-u_1^2)+g_{_{YB}}(g_{_B}+g_{_{YB}})(v_2^2-v_1^2)\nonumber\\
&&\qquad\;\quad\;+(-3g_{_2}^2+g_{_1}^2)(v_2^2-v_1^2)\Big]+\frac{v_2^2}{2}Y_u^\dagger Y_u,\nonumber\\
&&m_{uR}=m_{\tilde u}^2+\frac{1}{24}\Big[g_{_B}(g_{_B}+4g_{_{YB}})(u _1^2-u_2^2)+g_{_{YB}}(g_{_B}+4g_{_{YB}})(v_1^2-v_2^2)\nonumber\\
&&\qquad\;\quad\;+4g_{_1}^2(v_1^2-v_2^2)\Big]+\frac{v_1^2}{2}Y_u^\dagger Y_u,\label{massu}
\end{eqnarray}
which can be diagonalized by $Z^{\tilde u}$.

On the basis (${\rm Re}H_1^1$, ${\rm Re}H_2^2$, ${\rm Re}\tilde{\eta}_1$, ${\rm Re}\tilde{\eta}_2$),
the mass matrix for scalar Higgs bosons at the tree level reads
\begin{eqnarray}
&&M_h^2=u^2\times\nonumber\\
&&\left(\begin{array}{*{20}{c}}
{\frac{1}{4}\frac{g^2 x^2}{1+\tan\beta^2}+n^2\tan\beta}&{-\frac{1}{4}g^2\frac{x^2\tan\beta}{1+\tan^2\beta}}-n^2&
{\frac{1}{2}g_{_B}g_{_{YB}}\frac{x}{T}}&
{-\frac{1}{2}g_{_B}g_{_{YB}}\frac{x\tan\beta'}{T}}\\ [6pt]
{-\frac{1}{4}g^2\frac{ x^2\tan\beta}{1+\tan^2\beta}}-n^2&{\frac{1}{4}\frac{g^2\tan^2\beta x^2}{1+\tan\beta^2}+\frac{n^2}{\tan\beta}}&
{\frac{1}{2}g_{_B}g_{_{YB}}\frac{x\tan\beta}{T}}&{\frac{1}{2}g_{_B}g_{_{YB}}\frac{x\tan\beta\tan\beta'}{T}}\\ [6pt]
{\frac{1}{2}g_{_B}g_{_{YB}}\frac{x}{T}}&{\frac{1}{2}g_{_B}g_{_{YB}}\frac{x\tan\beta}{T}}&{\frac{g_{_B}^2}{1+\tan^2\beta'}+\tan\beta'N^2}&
{-g_{_B}^2\frac{\tan\beta'}{1+\tan^2\beta'}-N^2}\\ [6pt]
{-\frac{1}{2}g_{_B}g_{_{YB}}\frac{x\tan\beta'}{T}}&{\frac{1}{2}g_{_B}g_{_{YB}}\frac{x\tan\beta\tan\beta'}{T}}&
{-g_{_B}^2\frac{\tan\beta'}{1+\tan^2\beta^{'}}-N^2}&{g_{_B}^2\frac{\tan^2\beta'}{1+\tan^2\beta'}+\frac{N^2}{\tan\beta'}}
\end{array}\right)
\end{eqnarray}
where $g^2=g_{_1}^2+g_{_2}^2+g_{_{YB}}^2$, $T=\sqrt{1+\tan^2\beta}\sqrt{1+\tan^2\beta'}$,
$n^2=\frac{{\rm Re}B\mu}{u^2}$ and $N^2=\frac{{\rm Re}B\mu^{'}}{u^2}$, respectively. The matrix can be diagonalized by $Z^h$.

On the basis $(\tilde\lambda_B, \tilde\lambda_W, \tilde\lambda_{H_1^0}, \tilde\lambda_{H_2^0}, \tilde\lambda_{B'}, \tilde\lambda_{\eta_1}, \tilde\lambda_{\eta_2})$, new definition of the mass matrix for neutralinos is given by
\begin{eqnarray}
&&m_{\chi^0}^2=
\left(\begin{array}{ccccccc}M_1,&0,&-\frac{1}{2}g_1v_1,&\frac{1}{2}g_1v_2,&M_{BB'},&0,&0\\
0,&M_2,&\frac{1}{2}g_2v_1,&-\frac{1}{2}g_2v_2,&0,&0,&0\\
-\frac{1}{2}g_1v_1,&\frac{1}{2}g_2v_1,&0,&-\mu,&-\frac{1}{2}g_{_{YB}}v_1,&0,&0\\
\frac{1}{2}g_1v_2,&-\frac{1}{2}g_2v_2,&-\mu,&0,&\frac{1}{2}g_{_{YB}}v_2,&0,&0\\
M_{BB'},&0,&-\frac{1}{2}g_{_{YB}}v_1,&\frac{1}{2}g_{_{YB}}v_2,&M_B,&-g_{_B}u_1,&g_{_B}u_2\\
0,&0,&0,&0,&-g_{_B}u_1,&0,&-\mu'\\
0,&0,&0,&0,&g_{_B}u_2,&-\mu',&0
\end{array}\right).
\end{eqnarray}

\section{Constants $C_{abc}^{L,R}$ appeared in our calculation. \label{constants}}
\begin{eqnarray}
&&C_{\bar{\tilde g}\tilde q_i q_j}^L=-\sqrt2 g_3 Z^{\tilde q}_{i,j} e^{i\theta_3},\;\;
C_{\bar{\tilde g}\tilde q_i q_j}^R=\sqrt2 g_3 Z^{\tilde q}_{i,j+3} e^{-i\theta_3},\\
&&C_{\bar{q}_j\tilde q_i \tilde g}^L=\sqrt2 g_3 Z^{\tilde q *}_{i,j+3} e^{i\theta_3},\;\;
C_{\bar{q}_j\tilde q_i \tilde g}^R=-\sqrt2 g_3 Z^{\tilde q *}_{i,j} e^{-i\theta_3},\\
&&C_{\bar{\chi}_k^0 \tilde u_i u_j}^L=-\frac{1}{6}\Big(\sqrt2 g_1 Z^{N*}_{k,1}Z^{\tilde u}_{i,j}+3\sqrt2 g_2 Z^{N*}_{k,2}Z^{\tilde u}_{i,j}+\sqrt2 (g_{_{B}}+g_{_{YB}})Z^{N*}_{k,5}Z^{\tilde u}_{i,j}\nonumber\\
&&\qquad\qquad\;\;+6Y_{u,j}Z^{N*}_{k,4}Z^{\tilde u}_{i,j+3}\Big),\nonumber\\
&&C_{\bar{\chi}_k^0 \tilde u_i u_j}^R=\frac{1}{6}\Big(4\sqrt2 Z^{N}_{k,1}Z^{\tilde u}_{i,j+3}+\sqrt2 (g_{_{B}}+4g_{_{YB}})Z^{N}_{k,5}Z^{\tilde u}_{i,j+3}-6Y_{u,j}Z^{N}_{k,4}Z^{\tilde u}_{i,j}\Big),\\
&&C_{\bar{\chi}_k^0 \tilde d_i d_j}^L=-\frac{1}{6}\Big(\sqrt2 g_1 Z^{N*}_{k,1}Z^{\tilde d}_{i,j}-3\sqrt2 g_2 Z^{N*}_{k,2}Z^{\tilde d}_{i,j}+\sqrt2 (g_{_{B}}+g_{_{YB}})Z^{N*}_{k,5}Z^{\tilde d}_{i,j}\nonumber\\
&&\qquad\qquad\;\;+6Y_{d,j}Z^{N*}_{k,4}Z^{\tilde d}_{i,j+3}\Big),\nonumber\\
&&C_{\bar{\chi}_k^0 \tilde d_i d_j}^R=\frac{1}{6}\Big(-2\sqrt2 Z^{N}_{k,1}Z^{\tilde d}_{i,j+3}+\sqrt2 (g_{_{B}}-2g_{_{YB}})Z^{N}_{k,5}Z^{\tilde d}_{i,j+3}-6Y_{d,j}Z^{N}_{k,4}Z^{\tilde d}_{i,j}\Big),\\
&&C_{\bar{\chi}_k^+ \tilde u_i d_j}^L=\sum_{a=1}^3 \Big(-g_2 V_{k,1}^*Z^{CKM}_{a,j} Z^{\tilde u}_{i,a}+V_{k,2}^* Y_{u,a}Z^{CKM}_{a,j} Z^{\tilde u}_{i,a+3}\Big),\nonumber\\
&&C_{\bar{\chi}_k^+ \tilde u_i d_j}^R=U_{k,2}\sum_{a=1}^3Y_{d,a}Z^{CKM}_{a,j} Z^{\tilde u}_{i,a},\\
&&C_{\bar d_jd \tilde u_i {\chi}_k^-}^L=U_{k,2}^*\sum_{a=1}^3Y_{d,a}Z^{CKM*}_{a,j} Z^{\tilde u*}_{i,a},\nonumber\\
&&C_{\bar d_jd \tilde u_i {\chi}_k^-}^R=\sum_{a=1}^3 \Big(-g_2 V_{k,1}Z^{CKM*}_{a,j} Z^{\tilde u*}_{i,a}+V_{k,2} Y_{u,a}Z^{CKM*}_{a,j} Z^{\tilde u*}_{i,a+3}\Big),\\
&&C_{\bar{\chi}_k^- \tilde d_i u_j}^L=\sum_{a=1}^3 \Big(-g_2 U_{k,1}^*Z^{CKM*}_{j,a} Z^{\tilde d}_{i,a}+U_{k,2}^* Y_{d,a}Z^{CKM*}_{j,a} Z^{\tilde d}_{i,a+3}\Big),\nonumber\\
&&C_{\bar{\chi}_k^- \tilde d_i u_j}^R=V_{k,2}\sum_{a=1}^3Y_{u,a}Z^{CKM}_{j,a} Z^{\tilde d}_{i,a},\\
&&C_{\bar u_j \tilde d_i {\chi}_k^+}^L=V_{k,2}^*\sum_{a=1}^3Y_{u,a}Z^{CKM}_{j,a} Z^{\tilde d}_{i,a},\nonumber\\
&&C_{\bar u_j \tilde d_i {\chi}_k^+}^R=\sum_{a=1}^3 \Big(-g_2 U_{k,1}Z^{CKM}_{j,a} Z^{\tilde d}_{i,a}+U_{k,2} Y_{d,a}Z^{CKM}_{j,a} Z^{\tilde d}_{i,a+3}\Big),\\
&&C_{\bar u_i h_k u_i}=-\frac{1}{\sqrt2}Y_{u_i}Z^h_{k,2},\\
&&C_{\bar d_i h_k d_i}=-\frac{1}{\sqrt2}Y_{d_i}Z^h_{k,1}.
\end{eqnarray}

\end{document}